\documentclass[11pt]{article}   
\usepackage{booktabs,caption}
\usepackage[flushleft]{threeparttable}
\usepackage{amssymb}   
\usepackage{amsthm}    
\usepackage{color}
\usepackage{amsmath}    
\usepackage{amsfonts}
\usepackage{graphicx}
\usepackage{natbib}
\usepackage{dsfont}
\usepackage{setspace}
\usepackage{hyperref}
\usepackage{url}
\usepackage{pdflscape}
\usepackage{xspace}
\usepackage[margin=1.25in]{geometry}
\usepackage[gen]{eurosym}
\usepackage{epstopdf}  
\usepackage{fontenc}
\usepackage{placeins}
\usepackage{indentfirst}
\usepackage{enumitem}
\usepackage{multirow}
\usepackage{algorithm}
\usepackage{algpseudocode}

\theoremstyle{plain}
\newtheorem{theo}{Theorem}[section]
\newtheorem{coll}[theo]{Corollary}

\newtheorem{prop}[theo]{Proposition}
\newtheorem{assumption}[theo]{Assumption}

\theoremstyle{definition}

\theoremstyle{remark}
\newtheorem{remark}[theo]{Remark}

\bibpunct[, ]{(}{)}{,}{a}{}{,}

\def\1{\mathds{1}}

\renewcommand{\le}{\leqslant}
\renewcommand{\ge}{\geqslant}
\renewcommand{\leq}{\leqslant}

\newcommand{\be}{\begin{equation}}
\newcommand{\ee}{\end{equation}}
\newcommand{\bea}{\begin{eqnarray}}
\newcommand{\eea}{\end{eqnarray}}
\newcommand{\bean}{\begin{eqnarray*}}
\newcommand{\eean}{\end{eqnarray*}}
\newcommand{\vesp}{\vspace{0.15in}}

\begin{document}

\title{\bf Feasible sub-replicating portfolios for basket options via the block rearrangement algorithm}
\author{Mohamad Hassan Abou Daya\thanks{Mohamad Hassan Abou Daya, Department of Economics and Political Sciences at Vrije Universiteit Brussel (email: \texttt{mohamad-hassan.abou-daya@grenoble-em.com}).},\\ Carole Bernard\thanks{Carole Bernard, Department of Accounting, Law and Finance, Grenoble Ecole de Management, and Department of Economics and Political Sciences at Vrije Universiteit Brussel (email: \texttt{carole.bernard@grenoble-em.com}).}\ \ and Steven Vanduffel\thanks{Steven Vanduffel, Department of Economics and Political Sciences at Vrije Universiteit Brussel (email: \texttt{steven.vanduffel@vub.be}).}
}

\maketitle
\begin{abstract}
	We construct sub-replicating portfolios of vanilla options for a basket call when the marginal distributions of the underlying assets are known and their dependence is not. To do so, we reduce the $n$-asset problem to $n$ two-asset problems of the type solved by \cite{hobson2005sub}, and use the Block Rearrangement Algorithm\footnote{the BRA is a generalisation of the original rearrangement algorithm (RA) developed in  \cite{puccetti2012computation}.} (BRA) to approximate the minimising dependence among the assets. The cash component of the portfolio is computed exactly, so that its price is a valid lower bound, and reassembling the BRA output with the original marginals gives an upper bound computed by quadrature; together they bracket the minimum basket-option price. In log-normal examples with three and six assets the bracket has a relative width of at most $0.1\%$, and the corresponding sub-replicating portfolios are sparse, using only two strikes per asset, while the resulting bounds are tighter than those obtained from a general-purpose multi-marginal optimal-transport algorithm with strikes on a grid.
    We also give the lower bound and an optimal portfolio in closed form in three special cases: a deep in-the-money basket, a basket with one dominant asset, and jointly mixable marginals.
\end{abstract}

\vesp
\noindent Keywords:  Basket options, arbitrage-free bounds, model-free dependence, sub-replicating portfolios, Block Rearrangement Algorithm.

\vesp
\noindent JEL classification: G13, C63.

\section{Introduction}

Over the past two decades there has been increasing interest in finding price bounds for basket options. While the upper bound of the basket option price and the construction of super-replicating portfolios are by now well understood \citep{hobson2005sup,d2006static,pena2010upper,neufeld2023}, much less is known about the lower bound and the corresponding sub-replicating portfolios. In this paper we consider a basket of more than two assets with known marginal distributions, dependence uncertainty, and a single-period (static-arbitrage) setting.

To date, most progress on the lower bound of basket options has been made via semi-infinite linear programming. \cite{d2006static} obtain the lower bound in closed form when a single option price per asset is known, and a linear-programming lower bound, not shown to be sharp, when the forward prices are known as well. \cite{pena2010} give tractable linear programmes for the lower bound when the basket contains two assets, or when a forward price and a single call price per asset are known for any number of assets; \cite{pena2010upper} and \cite{cho2016computing} compute lower bounds numerically in the general case, by Dantzig--Wolfe decomposition and by discretising the semi-infinite programme respectively. \cite{neufeld2023} develop cutting-plane algorithms, in the spirit of the discretisation method of \cite{daum2011novel} but for a more general class of multi-asset payoffs, to compute bounds on basket options and other multi-asset derivatives, and report an experiment with 60 assets and 400 traded instruments. All of these methods take as input finitely many option prices rather than the full marginal distributions, i.e.\ a continuum of option prices, which is one source of difficulty in pinning down the lower bound. Another source is that the dependence structure attaining the lower bound is not known in closed form for more than two assets. When two assets are present the optimal dependence structure is the lower Fr\'echet copula \citep{dhaene1996dependency}; \cite{hobson2005sub} exploit this structure to construct sheep-track portfolios (STP) which optimally sub-replicate the basket option in the two-asset case. Their result is the starting point for the algorithm developed below, which uses STP to sub-replicate basket options containing any number of assets.

The work closest to ours is the multi-marginal optimal-transport algorithm of \cite{neufeld2022mmot}.
They replace each marginal constraint by finitely many moment constraints on a grid, solve the resulting linear semi-infinite programme, and recover a coupling of the original marginals by an operation that they call reassembly.
Applied to a basket call, this yields lower and upper bounds on the
minimum price, with a gap that vanishes as the grid is refined, together with a
sub-replicating portfolio of vanilla options whose strikes lie on the grid.
Our construction instead selects a small number of strikes per asset from the structure of the two-asset problem, yielding sub-replicating portfolios that provide lower bounds on the minimum basket-option price, with closed-form solutions in the special cases of Section~\ref{S2}. To obtain an upper bound on the minimum price, we apply their reassembly procedure to the output of the Block Rearrangement Algorithm (BRA), thereby constructing a coupling with the original marginals (Section~\ref{ssec:exact}). A key distinction is that the pair decomposition approach we use identifies suitable strike locations directly and does not rely on a prescribed grid. This makes it possible to
construct much sparser portfolios without sacrificing
accuracy.
Sections~\ref{ssec:nx} 
and~\ref{ssec:nx-marginals} compare the two methods on our examples and on marginals taken from their paper. With two strikes per asset our lower bounds are higher than theirs with $32$ to $256$ grid strikes per asset and our upper bounds are considerably tighter. Moreover, adding grid strikes to ours does not improve our lower bound. This shows that the location of the
strikes, and not their number, is crucial for obtaining tight bounds. Their algorithm, however, handles general cost functions and many more assets than our construction does. The two approaches are therefore complementary.

In a different direction, optimal-transport methods have been used to derive bounds on multi-asset derivatives. \cite{bartl2022marginal} formulate the dual transport problem when additional information on the dependence is available, and \cite{ansari2024improved} extract such information from the market prices of traded multi-asset options. Others have used neural networks and stochastic-gradient-descent algorithms to compute bounds of multi-asset derivatives in the presence of additional information \citep{aquino2020bounds,eckstein2020robust,eckstein2021robust,dragazi2024improved}. In general, bounds on multi-asset derivatives are tighter when additional information on the dependence structure is present.

We make four main contributions. First, we use the BRA to approximate the lower bound of the basket call option price and present a simple algorithm to construct feasible sub-replicating portfolios for basket options. By exploiting a decomposition into
two-asset problems, our method selects the strikes in a structured manner (rather than from a predetermined grid), yielding sparse portfolios. Second, we show that the cash component of the sub-replicating portfolio can be computed exactly by exploiting the piecewise-linear structure of the payoff function, so that no Monte-Carlo approximation of the infimum is needed; this exact computation also turns the framework from a feasibility argument into a verifiable optimality argument, in the sense that whenever the expected cost of the hedging portfolio coincides with the basket value under the BRA dependence structure, weak duality implies that both the dependence structure and the portfolio are optimal for the marginals used in the computation (Corollary~\ref{coll:optimality}). Third, we remove the restriction to the discretised marginals on which the rearrangement algorithms operate: the price of the portfolio under the original marginals and the basket value of the BRA output reassembled with the original marginals, in the sense of \cite{neufeld2022mmot}, bracket the true lower bound (Corollary~\ref{coll:bracket}), and in our examples the bracket has a relative width of at most $0.1\%$. Fourth, we provide closed-form expressions for the lower bound, together with optimal sub-replicating portfolios of vanilla options, in three special cases; the case of one dominant asset is characterised by a verifiable condition on the marginals alone. Our results contribute to the open question of characterising lower bounds of basket option prices in arbitrary dimensions.

The key idea behind our method is to exploit the two-asset sub-replication results of \cite{hobson2005sub} to construct feasible portfolios for baskets with any number of assets.
For a basket of $n$ assets, each asset $X_i$ is paired with the aggregate $Y_i=\sum_{j\ne i}X_j$ of the remaining assets, reducing the problem to $n$ instances of the two-asset sub-replication problem solved by \cite{hobson2005sub}. From each two-asset pair we extract strikes and construct the corresponding piecewise-linear payoff functions $f_i(x_i)$, and a cash component $\overline\lambda$ is added to enforce pointwise feasibility. While the pair decomposition applies in principle to any dimension, two practical difficulties arise. First, in dimension two the dependence structure minimising the sum in convex order, and hence every basket call price, is the lower Fr\'echet copula (the antimonotone dependence structure), but in dimension three and above no such closed-form structure exists \citep{bernard2014risk}; we therefore rely on the BRA to approximate it numerically. Second, computational cost and accuracy are sensitive to the number of assets, as we discuss in Section~\ref{S3}.

The rest of the paper is organised as follows. Section~\ref{S1} formulates the problem and presents the feasible solution. Section~\ref{S2} presents the optimal sub-replicating portfolio in three special cases. Section~\ref{sec:optimality} discusses why optimality of the construction cannot be guaranteed in advance and how it is verified after the fact. Section~\ref{S3} provides numerical examples and assesses their accuracy. Section~\ref{sc:Conclusion} concludes.

\section{Problem formulation and feasible solution\label{S1}}

Let $(\Omega,\mathcal{F},\mathbb{Q})$ be a probability space supporting non-negative random variables $X_1,\ldots,X_n$ representing the time-$T$ prices of $n$ underlying assets, and denote by $C_{X_i}(k):=\mathbb{E}_{\mathbb{Q}}[(X_i-k)^+]$ the call price function on $X_i$ at strike $k\ge 0$. Let $K$ be the basket call strike. Throughout the paper we work under a risk-neutral measure $\mathbb{Q}$, assume zero interest rates, no dividends, and unit weights in the basket; the extension to non-unit weights and constant interest rates follows by a trivial scaling argument \citep{laurence2005}. We use the same maturity for all options and suppress it from the notation.

We make the following standing assumption.

\begin{assumption}\label{ass:marginals}
For each $i=1,\ldots,n$, the marginal law of $X_i$ under $\mathbb{Q}$ admits a density on $(0,\infty)$ with full support and finite mean. The call price function $C_{X_i}(k)$ is therefore strictly positive for every $k\ge 0$, strictly decreasing, strictly convex on $[0,\infty)$, and continuously differentiable in $k$, with derivative $\partial_k C_{X_i}(k)=-\mathbb{Q}(X_i> k)=-\mathbb{Q}(X_i\ge k)$ (the two coincide because $\{X_i=k\}$ has measure zero under the density). 
\end{assumption}

\begin{remark}\label{rem:Y-discrete}
The construction below also uses call price functions $C_{Y_i}$ where $Y_i$ is a sum of marginals under the BRA-approximated dependence structure $\hat\mu$. By construction $\hat\mu$ is supported on a finite sample, so the law of $Y_i$ under $\hat\mu$ is discrete and $C_{Y_i}$ is continuous and convex but only \emph{piecewise} linear, with kinks at the atoms of $Y_i$. We therefore work with the right and left derivatives $C^{'+}_{Y_i}$ and $C^{'-}_{Y_i}$ in the auxiliary functions \eqref{phi1}--\eqref{phi3}.
\end{remark}

Although the method below applies, in principle, to any number of assets, we present the problem and its feasible solution for $n=3$ for ease of exposition; the extension to general $n$ is recorded in Remark~\ref{rem:general-n}.

The primal problem is the optimal-transport problem
\begin{equation}\label{eq:primal}
\inf_{\mu\in\Pi}\int_{\mathbb{R}^3_+}(x_1+x_2+x_3-K)^+\,\mu(dx_1,dx_2,dx_3),
\end{equation}
where $\Pi$ is the set of probability measures on $\mathbb{R}^3_+$ with the prescribed marginals, i.e.\ for each $i$
$$
\int_{\mathbb{R}^3_+}(x_i-k_i)^+\,\mu(dx_1,dx_2,dx_3) = C_{X_i}(k_i)\quad\text{for all } k_i\ge 0,
$$
$$
\int_{\mathbb{R}^3_+}\mu(dx_1,dx_2,dx_3)=1.
$$

The associated linear-programming dual is
$$
\sup_{\lambda,\nu_1,\nu_2,\nu_3}\;\lambda+\sum_{i=1}^{3}\int_{\mathbb{R}_+}C_{X_i}(k_i)\,\nu_i(dk_i),
$$
subject to the pointwise sub-replication constraint
$$
\sum_{i=1}^{3}\int_{\mathbb{R}_+}(x_i-k_i)^+\,\nu_i(dk_i) + \lambda \;\le\; (x_1+x_2+x_3-K)^+\quad\forall\,(x_1,x_2,x_3)\in\mathbb{R}^3_+,
$$
where $\lambda\in\mathbb{R}$ and the $\nu_i$ are signed Borel measures on $\mathbb{R}_+$ representing the strike profile of static option positions. Problem \eqref{eq:primal} is a multi-marginal optimal-transport problem with the continuous cost $(x_1+x_2+x_3-K)^+\le x_1+x_2+x_3$, which is integrable under Assumption~\ref{ass:marginals}; strong duality between \eqref{eq:primal} and the dual over integrable functions $\varphi_i(x_i)$ satisfying $\sum_i\varphi_i(x_i)\le(\sum_i x_i-K)^+$ therefore holds by the classical duality theorem of \cite{kellerer1984duality} (see also \cite{bartl2022marginal} for the marginal-uncertainty setting and \cite{beiglbock2013model} for the martingale variant). The dual written above ranges over static call positions $\nu_i$ rather than over arbitrary integrable functions of $x_i$; every continuous piecewise-linear function of $x_i$ with finitely many kinks is of this form, and the portfolios constructed below are of this type, but we do not claim that restricting the dual to call positions preserves its value. Only weak duality, i.e.\ the inequality (expected cost of a pointwise sub-replicator) $\le$ (value of \eqref{eq:primal}), is used in the proofs, so the precise class of dual instruments does not affect the results below.
\subsection{Feasible sub-replicating portfolio}\label{ssec:feasible}
Let $\mu^*\in\Pi$ be a dependence structure attaining (or approximating) the infimum in \eqref{eq:primal}, and let $(X_1^*,X_2^*,X_3^*)$ be a random vector with joint law $\mu^*$. Define the auxiliary random variables
$$
Y_1^* := X_2^*+X_3^*,\qquad Y_2^* := X_1^*+X_3^*,\qquad Y_3^* := X_1^*+X_2^*,
$$
and write $\mu_{Y_i}^*$ for the marginal law of $Y_i^*$ under $\mu^*$, with corresponding call price function $C_{Y_i}(k):=\mathbb{E}_{\mu^*}[(Y_i^*-k)^+]$. For each pair $(X_i^*,Y_i^*)$ the basket option on $X_i^*+Y_i^*$ is a two-asset basket option, and the optimal sub-replicating portfolio for this two-asset problem is given by \cite{hobson2005sub}. In what follows, we use the measures derived from the two-asset case to construct a feasible portfolio for the three-asset basket option.

In practice, the exact minimiser $\mu^*$ is unknown in dimension three or higher (see the discussion in Section~\ref{ssec:algorithm}); we therefore use an approximate dependence structure $\hat\mu$ obtained from the BRA, and the marginals $\mu_{Y_i}^*$ above are replaced by their BRA-approximated counterparts $\hat\mu_{Y_i}$.

\noindent Define the auxiliary functions
\begin{equation}
\phi_1 \left( x_1\right) =C_{X_1}^{'+}\left( x_1\right) +C_{Y_1}^{'-}\left( K-x_1\right) +1
	\label{phi1}
\end{equation}

\begin{equation}
\phi_2 \left( x_2\right) =C_{X_2}^{'+}\left( x_2\right) +C_{Y_2}^{'-}\left( K-x_2\right) +1
	\label{phi2}
\end{equation}

\begin{equation}
\phi_3 \left( x_3\right) =C_{X_3}^{'+}\left( x_3\right) +C_{Y_3}^{'-}\left( K-x_3\right) +1
	\label{phi3}
\end{equation}
\noindent where $C^{'+}$ and $C^{'-}$ are respectively the right and left derivatives of the call option price function with respect to the strike.

\noindent Define 

$$
A_1 = \lbrace x_1:  \phi_1 \left( x_1\right)>0 \rbrace
$$

$$
A_2 = \lbrace x_2:  \phi_2 \left( x_2\right)>0 \rbrace
$$

$$
A_3 = \lbrace x_3:  \phi_3 \left( x_3\right)>0 \rbrace
$$

\noindent Let $\lbrace K_1^{2i}, K_1^{2i-1} \rbrace$, $\lbrace K_2^{2i}, K_2^{2i-1} \rbrace$, and $\lbrace K_3^{2i}, K_3^{2i-1} \rbrace$ be the boundaries of the naturally ordered intervals $\left({A_{1,i}}\right)_{1\leq i \leq n_1}$, $\left({A_{2,i}}\right)_{1\leq i \leq n_2}$, and $\left({A_{3,i}}\right)_{1\leq i \leq n_3}$ respectively. Let $K_{y_i}^j=K-  K_{i}^{2n_i-j+1}$. In other words, the $K_i^j$ are the boundary points of the set on which $\phi_i$ is positive; when $\phi_i$ is continuous these are zeros of $\phi_i$, with the convention that $K_i^1=0$ when $A_i$ contains a right-neighbourhood of $0$. Since $\phi_i(x)=-\mathbb{Q}(X_i>x)\le 0$ for $x\ge K$, one always has $A_i\subseteq[0,K)$. Throughout we assume that each $A_i$ is a finite union of intervals; this is automatic when the law of $Y_i$ is discrete, as it is for the BRA output (Remark~\ref{rem:Y-discrete}), and it holds in the examples of Section~\ref{S3}. 

\noindent Define
$$
\overline{\nu}_{j}\left( dk_{j}\right) =\delta_0\left( k_{j}\right) dk_{j}+\sum ^{2n_j}_{i=1}\left( -1\right) ^{i}\delta _{K^i_j}\left( k_{j}\right) dk_{j}\;\;\; ,j= 1,2,3
$$

$$
\overline{\nu}_{y_j}\left( dk_{y_j}\right) =\delta_0\left( k_{y_j}\right) dk_{y_j}+\sum ^{2n_j}_{i=1}\left( -1\right) ^{i}\delta _{K^i_{y_j}}\left( k_{y_j}\right) dk_{y_j}\;\;\; ,j= 1,2,3
$$

$$
{\lambda_j }=\sum ^{n_j}_{i=1}\left(K_{j}^{2i}-K_{j}^{2i-1}\right)-K=\sum ^{n_j}_{i=1}\left(K_{y_j}^{2i}-K_{y_j}^{2i-1}\right)-K\;\;\; ,j= 1,2,3
$$

\noindent From the result of \cite{hobson2005sub}, each triplet $(\lambda_j,\overline{\nu}_{j},\overline{\nu}_{y_j})$ is a feasible sub-replicating portfolio for the two-asset basket $X_j+Y_j$ if options on $Y_j$ are traded, and it is optimal when $Y_j$ has a continuous law, so that $\phi_j$ is continuous and vanishes at the endpoints of $A_j$; for the discrete BRA-approximated $Y_j$ of Remark~\ref{rem:Y-discrete} optimality of the two-asset triplet is not asserted. The construction below uses only the strikes of the $X_j$-legs, so neither property is needed for the results that follow. The constant $\lambda_j$ defined above is the cash component of the \emph{two-asset} Hobson--Laurence--Wang sub-replicating portfolio for the pair $(X_j,Y_j)$; it should not be confused with the cash component $\overline{\lambda}$ defined in \eqref{eq:lambda-def} for the \emph{combined three-asset} portfolio constructed below. Of each triplet we retain only the $X_j$-leg $\overline{\nu}_{j}$, and we combine these legs into the portfolio

$$
f_{1}\left( x_1\right) + f_{2}\left( x_2\right) +f_{3}\left( x_3\right) +\overline{\lambda }
$$

\noindent where 

\begin{equation}\label{eq:fi}
f_{i}\left( z\right) =z^{+}+\sum ^{n_i}_{a=1}\left\{ \left( z-K_i^{2a}\right) ^{+}-\left( z-K_i^{2a-1}\right) ^{+}\right\}
\end{equation}

\noindent Feasibility amounts to establishing the pointwise sub-replication inequality
\begin{equation}\label{eq:subrep}
f_{1}(x_1) + f_{2}(x_2) + f_{3}(x_3) + \overline{\lambda } \;\le\; (x_1+x_2+x_3-K)^+\qquad\forall\,(x_1,x_2,x_3)\in\mathbb{R}^3_+.
\end{equation}
By construction each $f_i$ is a piecewise-linear function whose slope alternates between $0$ and $1$, with $f_i(0)=0$, $0\le f_i(z)\le z$ on $\mathbb{R}_+$, and $f_i(z)=z-L_i$ for $z\ge K_i^{2n_i}$, where
$$
L_i := \sum_{a=1}^{n_i}\bigl(K_i^{2a}-K_i^{2a-1}\bigr)\ge 0.
$$
The function
$$
\Psi(x_1,x_2,x_3) := (x_1+x_2+x_3-K)^+ - \bigl(f_1(x_1)+f_2(x_2)+f_3(x_3)\bigr)
$$
is therefore piecewise linear on a finite polyhedral subdivision of $\mathbb{R}^3_+$. It is bounded below: on the region $\{x_1+x_2+x_3\le K\}$ one has $\Psi\ge -\sum f_i(x_i)\ge -K$, and on $\{x_1+x_2+x_3>K\}$ one has $\Psi=\sum(x_i-f_i(x_i))-K\ge -K$. Hence
\begin{equation}\label{eq:lambda-def}
\overline{\lambda } := \inf_{(x_1,x_2,x_3)\in\mathbb{R}^3_+}\Psi(x_1,x_2,x_3)\;\in\;[-K,0]
\end{equation}
exists and is finite. With this choice of $\overline{\lambda}$, the inequality \eqref{eq:subrep} holds by definition.

Because $\Psi$ is piecewise linear, its infimum on $\mathbb{R}^n_+$ is attained at a vertex of the underlying polyhedral subdivision; these vertices can be enumerated explicitly, yielding an exact formula for $\overline{\lambda}$.

\begin{prop}[Exact computation of $\overline{\lambda}$]\label{prop:exact-lambda}
Let the strikes of each asset be ordered, $0\le K_i^1\le K_i^2\le\cdots\le K_i^{2n_i}$, so that each $f_i$ in \eqref{eq:fi} has slopes in $\{0,1\}$ (this is the case for the strikes of Section~\ref{ssec:feasible}, and it is preserved by the strike-refinement step of Algorithm~\ref{alg:subrep}). For each $i=1,\ldots,n$, let
$$
\mathcal{B}_i := \{0\}\cup\{K_i^1,\ldots,K_i^{2n_i}\}\cup\{K\}
$$
be the set of breakpoints of $f_i$ together with the boundaries $0$ and $K$. Then
\begin{equation}\label{eq:lambda-exact}
\overline{\lambda} = \min_{(x_1,\ldots,x_n)\,\in\,\mathcal{B}_1\times\cdots\times\mathcal{B}_n} \Psi(x_1,\ldots,x_n).
\end{equation}
\end{prop}

\begin{proof}
Let $\mathcal{H}$ be the finite family of hyperplanes $\{x_i=b\}$, $b\in\{0\}\cup\{K_i^1,\ldots,K_i^{2n_i}\}$, $i=1,\ldots,n$, together with the hyperplane $\{\sum_i x_i=K\}$. These hyperplanes cut $\mathbb{R}^n_+$ into finitely many closed polyhedral cells, and $\Psi$ is affine on each cell, because each $f_i$ is affine between consecutive breakpoints and the basket payoff is affine on either side of $\{\sum_i x_i=K\}$. Every cell is contained in $\mathbb{R}^n_+$, hence contains no line, and $\Psi\ge -K$ on it. By the fundamental theorem of linear programming, an affine function that is bounded below on a non-empty polyhedron containing no line attains its minimum at a vertex of that polyhedron. Since the cells cover $\mathbb{R}^n_+$, the infimum $\overline{\lambda}$ is attained at a vertex of one of the cells; we write $\mathcal{V}$ for the finite set of all such vertices, so that $\overline{\lambda}=\min_{v\in\mathcal{V}}\Psi(v)$.

A vertex $v\in\mathcal{V}$ is the intersection of $n$ linearly independent hyperplanes of $\mathcal{H}$. Two cases arise. If all of them are of the form $\{x_i=b\}$, then by independence each coordinate is fixed by exactly one of them and $v\in\mathcal{B}_1\times\cdots\times\mathcal{B}_n$. Otherwise one of them is $\{\sum_i x_i=K\}$ and the other $n-1$ fix $n-1$ distinct coordinates, so there is exactly one index $m$ with
$$
v_m = K-\sum_{i\ne m}v_i\in[0,K],\qquad v_i\in\mathcal{B}_i\quad(i\ne m).
$$
If $v_m\in\mathcal{B}_m$ we are done. Otherwise $v_m$ is not a breakpoint of $f_m$, so $f_m$ is affine on a neighbourhood of $v_m$ with slope $s\in\{0,1\}$, and we slide the $m$-th coordinate. Put $g(t):=\Psi(v+te_m)$ for $t\ge -v_m$; since $\sum_i v_i=K$,
$$
g(t)\;=\;t^+-f_m(v_m+t)+c,\qquad c:=-\sum_{i\ne m}f_i(v_i).
$$
If $s=1$, let $b$ be the smallest element of $\mathcal{B}_m$ larger than $v_m$; it exists because $K\in\mathcal{B}_m$ and $v_m<K$ (recall $v_m\notin\mathcal{B}_m$). No breakpoint of $f_m$ lies in $(v_m,b)$, so $f_m$ has slope $1$ on $[v_m,b]$ and $g(t)=t-\bigl(f_m(v_m)+t\bigr)+c=g(0)$ for $0\le t\le b-v_m$. If $s=0$, let $b$ be the largest element of $\mathcal{B}_m$ smaller than $v_m$; it exists because $0\in\mathcal{B}_m$ and $v_m>0$. Then $f_m$ is constant on $[b,v_m]$ and $g(t)=-f_m(v_m)+c=g(0)$ for $b-v_m\le t\le 0$. In both cases the point $v'$ obtained from $v$ by replacing $v_m$ with $b$ lies in $\mathcal{B}_1\times\cdots\times\mathcal{B}_n$ and satisfies $\Psi(v')=\Psi(v)$. (The case $s=1$ with no breakpoint above $v_m$ is the only place where the point $K\in\mathcal{B}_m$ is needed. The ordering of the strikes enters only through $s\in\{0,1\}$ and through the lower bound $\Psi\ge-K$. For an arbitrary finite family of strikes the slopes of $f_m$ are still integers and $\Psi$ is still bounded below, since every unbounded cell lies in $\{\sum_i x_i\ge K\}$ and $\Psi$ has slope $0$ along each of its recession directions; the same sliding then gives $\Psi(v')\le\Psi(v)$, because $g$ is non-decreasing on $[b-v_m,0]$ when $s\le 0$ and non-increasing on $[0,b-v_m]$ when $s\ge 1$. Hence \eqref{eq:lambda-exact} holds without the ordering assumption.)

Consequently $\overline{\lambda}=\min_{v\in\mathcal{V}}\Psi(v)\ge\min_{x\in\mathcal{B}_1\times\cdots\times\mathcal{B}_n}\Psi(x)$, while the reverse inequality is trivial because $\mathcal{B}_1\times\cdots\times\mathcal{B}_n\subset\mathbb{R}^n_+$. This proves \eqref{eq:lambda-exact}.
\end{proof}

The number of evaluations in \eqref{eq:lambda-exact} is $\prod_{i=1}^{n}|\mathcal{B}_i| \le \prod_{i=1}^{n}(2n_i+2)$, with equality when $0$, $K$ and the strikes of each asset are pairwise distinct. When each asset has a single pair of strikes ($n_i=1$), this equals $4^n$; for the three-asset example ($n=3$) this is $64$ evaluations, and for the six-asset example ($n=6$) it is $4{,}096$. Even for nine assets, $4^9 = 262{,}144$ evaluations are computed in under a second on a standard machine. The exact computation of $\overline{\lambda}$ therefore eliminates the Monte-Carlo approximation error entirely, at negligible computational cost for the basket sizes considered in this paper.

We summarise the resulting construction as follows.

\begin{theo}[Feasible sub-replicating portfolio]\label{thm:feasible}
Under Assumption~\ref{ass:marginals}, let $(X_i^*,Y_i^*)$ be the pairs constructed in Section~\ref{ssec:feasible} from a dependence structure $\mu^*\in\Pi$, let $f_1,f_2,f_3$ be the piecewise-linear functions defined in \eqref{eq:fi} and let $\overline{\lambda}$ be defined by \eqref{eq:lambda-def}. Then the portfolio
$$
\Pi^{\mathrm{sub}}(x_1,x_2,x_3) := f_1(x_1)+f_2(x_2)+f_3(x_3)+\overline{\lambda}
$$
is a pointwise sub-replicator of the basket call payoff on $\mathbb{R}^3_+$. Its expected cost under any dependence structure $\mu\in\Pi$ is
$$
\mathbb{E}_{\mu}\!\left[\Pi^{\mathrm{sub}}\right] = \sum_{i=1}^{3}\mathbb{E}_{\mu}[f_i(X_i)] + \overline{\lambda},
$$
which depends only on the marginals (and not on the joint law) since each $f_i$ is a function of $X_i$ alone.
\end{theo}

\begin{proof}
Pointwise sub-replication on $\mathbb{R}^3_+$ is immediate from the definition of $\overline{\lambda}$ in \eqref{eq:lambda-def}: by construction $\Pi^{\mathrm{sub}}(x)=(\sum_i x_i-K)^+ - \Psi(x) + \overline{\lambda}$, and $\overline{\lambda}\le \Psi(x)$ for every $x\in\mathbb{R}^3_+$, so $\Pi^{\mathrm{sub}}(x)\le (\sum_i x_i - K)^+$. The expected-cost identity follows from linearity of expectation. Each $f_i(X_i)$ is integrable since $0\le f_i(z)\le z$ on $\mathbb{R}_+$ (by the piecewise-linear structure of $f_i$) and $\mathbb{E}_\mu[X_i]<\infty$ by the finite-mean part of Assumption~\ref{ass:marginals}; the constant $\overline{\lambda}$ is finite by \eqref{eq:lambda-def}. Marginal-only dependence follows because each $f_i$ is a measurable function of $X_i$ alone, so $\mathbb{E}_\mu[f_i(X_i)]$ depends only on the law of $X_i$, which is fixed across $\mu\in\Pi$.
\end{proof}

\begin{remark}\label{rem:bra-vs-true}
When the dependence structure $\mu^*$ used to define $(X_i^*,Y_i^*)$ is replaced by the BRA approximation $\hat\mu$, Theorem~\ref{thm:feasible} continues to deliver a pointwise sub-replicating portfolio, but the strikes $\{K_i^j\}$ depend on $\hat\mu$. Consequently the expected cost of $\Pi^{\mathrm{sub}}$ approximates the lower bound $\inf_{\mu\in\Pi}\mathbb{E}_\mu[(X_1+X_2+X_3-K)^+]$ rather than attaining it exactly. Note that even with the exact minimiser $\mu^*$ the strikes inherited from the two-asset problems are not guaranteed to be dual optimal for the three-asset problem; see Section~\ref{sec:optimality} for a discussion of both sources of sub-optimality. Corollary~\ref{coll:optimality} below provides a verifiable sufficient condition under which the approximation is in fact exact for the marginals used in the computation, and the numerical examples in Section~\ref{S3} document the size of the gap in the regimes where it is not.
\end{remark}

\begin{coll}[Verification of optimality by weak duality]\label{coll:optimality}
Let $\mu_1,\ldots,\mu_n$ be marginal laws on $\mathbb{R}_+$ with finite means (for instance the discretised marginals used in the numerical implementation), let $\Pi$ be the set of dependence structures with these marginals, and let $\Pi^{\mathrm{sub}}=f_1(x_1)+\cdots+f_n(x_n)+\overline{\lambda}$ be the portfolio of Theorem~\ref{thm:feasible} built from an arbitrary family of ordered strikes $0\le K_i^1\le\cdots\le K_i^{2n_i}$ (not necessarily the zero-crossings of $\phi_i$), with $\overline{\lambda}$ defined by \eqref{eq:lambda-def} and computed exactly via Proposition~\ref{prop:exact-lambda}; the pointwise inequality \eqref{eq:subrep} then holds for any strikes and any marginals, since it uses only the definition of $\overline{\lambda}$. For any $\hat\mu\in\Pi$ (in particular, for the BRA output) define
$$
HV \;:=\; \sum_{i=1}^{n}\mathbb{E}_{\mu_i}[f_i(X_i)] + \overline{\lambda},\qquad
BV \;:=\; \mathbb{E}_{\hat\mu}\!\left[\Bigl(\sum_{i=1}^{n}X_i - K\Bigr)^{+}\right],
$$
noting that $HV$ depends only on the marginals (each $f_i$ is a function of $X_i$ alone and $\overline{\lambda}$ is a constant). If $HV = BV$, then
\begin{enumerate}[label=(\roman*)]
\item $\hat\mu$ attains the infimum: $BV \;=\; \inf_{\mu\in\Pi}\mathbb{E}_\mu[(\sum_{i=1}^{n}X_i - K)^+]$;
\item $\Pi^{\mathrm{sub}}$ is an optimal sub-replicating portfolio: its expected cost equals the lower bound.
\end{enumerate}
\end{coll}

\begin{proof}
For any $\mu\in\Pi$, pointwise sub-replication gives $\Pi^{\mathrm{sub}}(x)\le (\sum_i x_i-K)^+$ on $\mathbb{R}^n_+$, hence $\mathbb{E}_\mu[\Pi^{\mathrm{sub}}]\le \mathbb{E}_\mu[(\sum_i X_i - K)^+]$. Because each $f_i$ depends only on $X_i$ and $\overline{\lambda}$ is a constant, $\mathbb{E}_\mu[\Pi^{\mathrm{sub}}]=HV$ for every $\mu\in\Pi$. Taking the infimum over $\mu$ on the right yields
$$
HV \;\le\; \inf_{\mu\in\Pi}\mathbb{E}_\mu\!\left[\Bigl(\sum_{i=1}^{n}X_i - K\Bigr)^{+}\right] \;\le\; BV,
$$
where the second inequality holds because $\hat\mu\in\Pi$. Under the hypothesis $HV=BV$ all three quantities coincide, which is precisely (i) and (ii). Only weak duality has been used.
\end{proof}

\begin{remark}[Verifiability and discretisation]\label{rem:certificate}
The hypothesis of Corollary~\ref{coll:optimality} is fully numerical: $HV$ is a sum of marginal expectations and the exact constant $\overline{\lambda}$, while $BV$ is computed from the rearranged sample. The exact computation of $\overline{\lambda}$ via Proposition~\ref{prop:exact-lambda} removes the Monte-Carlo noise that would make an equality test impractical, and the $HV$-maximising strike refinement of Section~\ref{ssec:algorithm} is what closes the gap in the regimes we report. Two points deserve emphasis. First, the corollary uses weak duality only; no attainment or strong-duality result is needed. Second, in the implementation both $HV$ and $BV$ are computed on the discretised marginals $\hat\mu_i$ ($N$ equiprobable atoms), so the BRA output belongs to $\hat\Pi$, the set of dependence structures with these discretised marginals, and the certificate concerns $\inf_{\hat\Pi}$, the lower bound of the discretised problem. Because $\Pi^{\mathrm{sub}}$ is a pointwise sub-replicator on all of $\mathbb{R}^n_+$, its price under the original marginals, $HV^{\mathrm{ex}}:=\sum_i\mathbb{E}_{\mu_i}[f_i(X_i)]+\overline{\lambda}$ with $\mu_i$ the original laws, is a valid lower bound for the original problem \eqref{eq:primal}; it differs from $HV$ by the marginal discretisation error only. Section~\ref{ssec:exact} complements this lower bound with an upper bound for the original problem, so that the certificate is not confined to the discretised marginals.
\end{remark}

\begin{remark}[Generalisation to $n$ assets]\label{rem:general-n}
The construction extends to a basket of $n$ assets in a straightforward way. Apply BRA to obtain an approximate dependence structure $\hat\mu$ of the $n$ marginals supported on a sample $(x_1^{(m)},\ldots,x_n^{(m)})_{m=1}^M$ that approximately minimises $\mathbb{E}_\mu[(X_1+\cdots+X_n-K)^+]$ over $\mu\in\Pi$ (or, by default, the variance of the sum; see Section~\ref{ssec:algorithm}). For each asset $i$, define $Y_i^*:=\sum_{j\ne i}X_j^*$ under $\hat\mu$, form the pair $(X_i^*,Y_i^*)$, and apply the two-asset method of \cite{hobson2005sub} to obtain the strikes $\{K_i^j\}$ and the piecewise-linear payoff $f_i$. The cash component is then defined by
$$
\overline{\lambda} := \inf_{(x_1,\ldots,x_n)\in\mathbb{R}^n_+}\!\left\{\Bigl(\sum_{i=1}^{n}x_i - K\Bigr)^+ - \sum_{i=1}^{n}f_i(x_i)\right\}\in[-K,0],
$$
and computed exactly via Proposition~\ref{prop:exact-lambda}. The bounds $0\le f_i(x_i)\le x_i$ remain valid in any dimension, so $\overline{\lambda}$ is finite and the resulting portfolio is a pointwise sub-replicator of the $n$-asset basket call payoff.
\end{remark}

\subsection{Algorithm implementation}\label{ssec:algorithm}

The most important step in constructing the feasible portfolio is to identify the dependence structure of the marginals that minimises the basket option price. We use the BRA \citep[BRA hereafter;][]{bernard2016algorithms,boudt2018block}, a block-wise extension of the rearrangement algorithm of \cite{puccetti2012computation}. \cite{bernard2018rearrangement} provide numerical evidence that, among the dependence structures compatible with given marginals and a given distribution of the sum, the one returned by BRA has almost maximum entropy; in this sense BRA returns a typical dependence structure at the lower bound of the sum, not an exceptional one. More concretely, BRA seeks the dependence structure for which the sum of the marginals has the lowest variance, that is, the joint law under which the sum is as close to constant as the marginals allow. Minimising the variance is a proxy: the quantity of interest is $\mathbb{E}[(\sum_i X_i-K)^+]$, and for $n\ge 3$ the two objectives need not be minimised by the same dependence structure, since the sum has no convex-order minimum in general \citep{bernard2014risk}. This is why we also run the algorithms with the $K$-aware objective of Remark~\ref{rem:objectives} and select their output by the hedging value $HV$. We refer to \cite{bernard2018rearrangement} for a detailed exposition; the improved variant of \cite{bernard2026improved}, which selects the block sizes dynamically, can be substituted without changing the construction below.

We now explain why BRA is needed at all. When only two assets are present, the dependence structure minimising $X_1+X_2$ in convex order, and hence $\mathbb{E}[(X_1+X_2-K)^+]$ for every $K$, is the lower Fr\'echet copula (the antimonotone dependence structure), which is a valid copula in dimension two; \cite{hobson2005sub} exploit this known structure directly to construct optimal sub-replicating portfolios. In dimension three and above, however, the lower Fr\'echet bound $\max\bigl(\sum_{i=1}^{n}u_i - n + 1,\; 0\bigr)$ is no longer a valid copula \citep[Section~2.10]{nelsen2006introduction}, and no closed-form expression for the dependence structure minimising $\mathbb{E}[(\sum_{i=1}^n X_i-K)^+]$ is available in general. BRA provides a numerical approximation to this unknown optimal structure. Because BRA is a heuristic, it may converge to a local minimum rather than the global one; we mitigate this by running it multiple times, with random initialisations and with both objectives, and selecting the configuration with the highest hedging value $HV$ (see below).

The full procedure is summarised in Algorithm~\ref{alg:subrep}.

\begin{algorithm}[H]
\caption{Construction of the feasible sub-replicating portfolio}\label{alg:subrep}
\small
\begin{algorithmic}[1]
\Require Marginal laws $\mu_1,\ldots,\mu_n$; basket strike $K$; discretisation size $N$; number of rearrangement restarts $R$; stall window $W$
\Ensure Strikes $\{K_i^j\}$, piecewise-linear payoffs $\{f_i\}$, cash component $\overline{\lambda}$, hedging value $HV$, bounds $HV^{\mathrm{ex}}\le BV^{\mathrm{ex}}$
\For{$i=1,\ldots,n$}
    \State Discretise $\mu_i$ into $N$ equiprobable atoms $\{x_i^{(m)}\}_{m=1}^{N}$
\EndFor
\State $HV_{\mathrm{best}}\gets -\infty$
\For{$r=1,\ldots,R$}
    \State Apply RA or BRA with a fresh random initialisation to the matrix $[x_i^{(m)}]_{m,i}$, stopping after $W$ consecutive iterations without improvement \Comment{objective: $\mathrm{var}(\sum_i X_i)$ or the $K$-aware $\mathbb{E}[(\sum_i X_i - K)^+]$ of Remark~\ref{rem:objectives}}
    \State Let $\hat\mu^{(r)}$ denote the resulting dependence structure on the rearranged sample
    \For{$i=1,\ldots,n$}
        \State Form the empirical sample of $Y_i:=\sum_{j\ne i}X_j$ under $\hat\mu^{(r)}$
        \State Compute the empirical call price functions $C_{X_i}$, $C_{Y_i}$
        \State $\phi_i(x_i)\gets C_{X_i}^{'+}(x_i)+C_{Y_i}^{'-}(K-x_i)+1$
        \State $A_i\gets\{x_i\ge 0:\phi_i(x_i)>0\}$
        \State Extract the ordered interval boundaries $K_i^1<K_i^2<\cdots<K_i^{2n_i}$ of $A_i$
        \State If $K_i^{2n_i}$ exceeds the largest atom of $X_i$, set $K_i^{2n_i}\gets K$ \Comment{worthless call on the sample; this can only increase $\overline{\lambda}$}
        \State $f_i(z)\gets z+\sum_{a=1}^{n_i}\bigl[(z-K_i^{2a})^+-(z-K_i^{2a-1})^+\bigr]$
    \EndFor
    \If{$A_i=\emptyset$ for some $i$}
        \State \Return the linear portfolio $\Pi^{\mathrm{sub}}_1=\sum_{i=1}^{n}x_i-K$ of Proposition~\ref{prop:case1}, which is optimal \Comment{Case~1; no further search is needed}
    \EndIf
    \State Whether $0$ and $K$ belong to the strike set of asset $i$ depends on the sign of $\phi_i$ near the boundary, which is numerically ambiguous when $\phi_i$ is small there: resolve the $2^n$ boundary choices by an exact search, visiting them in decreasing order of $\sum_i\mathbb{E}[f_i(X_i)]$ and stopping once this quantity no longer exceeds the best $HV$ found \Comment{$\overline{\lambda}\le 0$, hence $HV\le\sum_i\mathbb{E}[f_i(X_i)]$}
    \State $\mathcal{B}_i\gets\{0\}\cup\{K_i^1,\ldots,K_i^{2n_i}\}\cup\{K\}$ for each $i$
    \State $\overline{\lambda}^{(r)}\gets\min\bigl\{\Psi(x):x\in\mathcal{B}_1\times\cdots\times\mathcal{B}_n\bigr\}$ \Comment{exact, by Proposition~\ref{prop:exact-lambda}}
    \State $HV^{(r)}\gets\sum_{i=1}^{n}\mathbb{E}_{\mu_i}[f_i(X_i)]+\overline{\lambda}^{(r)}$
    \If{$HV^{(r)}>HV_{\mathrm{best}}$}
        \State $HV_{\mathrm{best}}\gets HV^{(r)}$;\ \ store $\{K_i^j\},\{f_i\},\overline{\lambda}^{(r)}$ and $\hat\mu^{(r)}$
    \EndIf
\EndFor
\State \textit{Strike refinement:} starting from the stored strikes, run a derivative-free local search (Nelder--Mead alternated with Powell's method) on the map $\{K_i^j\}\mapsto HV(\{K_i^j\})$ (with the number of strikes per asset and their sign pattern fixed, the strikes of each asset kept sorted so that Proposition~\ref{prop:exact-lambda} applies, and strikes at $0$ or $K$ held fixed) until no further progress is made; recompute $\overline{\lambda}$ via Proposition~\ref{prop:exact-lambda} and update $HV_{\mathrm{best}}$; keep the refined portfolio if it is worth more than the strike-free portfolios of Remark~\ref{rem:trivial}
\State \textit{Original marginals (Section~\ref{ssec:exact}):} repeat the search with $HV^{\mathrm{ex}}$ as objective, with the strikes placed at $K$ kept there and, in a second run, released; compute $BV^{\mathrm{ex}}$ by reassembling the stored $\hat\mu^{(r)}$ \Comment{$HV^{\mathrm{ex}}\le\inf_\Pi\le BV^{\mathrm{ex}}$}
\State \Return $\{K_i^j\}$, $\{f_i\}$, $\overline{\lambda}$, $HV_{\mathrm{best}}$, $HV^{\mathrm{ex}}$, $BV^{\mathrm{ex}}$
\end{algorithmic}
\end{algorithm}

We have four practical recommendations. First, discretise the marginal distributions into at least $50{,}000$ atoms. Accurate strikes are essential for the cash component to be accurate; since the lower bound of the basket option is typically small, any inaccuracy in the strikes propagates directly to the portfolio value.

Second, run BRA at least ten times, since it occasionally converges to a poor local minimum, and let each run converge: with random block partitions, a stall rule of a few iterations without improvement stops BRA long before convergence, and we use a window $W$ of a few hundred iterations. Importantly, we recommend selecting the BRA output that maximises $HV = \sum_{i=1}^{n}\mathbb{E}[f_i(X_i)] + \overline{\lambda}$ rather than the output with the lowest variance of the sum. Two BRA outputs with similar variance can produce very different strike configurations, and the configuration with slightly higher variance may yield a tighter sub-replicating portfolio. Since $\overline{\lambda}$ is computed exactly via Proposition~\ref{prop:exact-lambda}, evaluating $HV$ for each BRA output incurs negligible additional cost.

Third, the Rearrangement Algorithm (RA) of \cite{puccetti2012computation} can complement BRA: running both algorithms with multiple random initialisations and selecting the output with the highest $HV$ across all runs further tightens the sub-replication gap.

Fourth, once candidate strikes $\{K_i^j\}$ have been identified from the zero-crossings of $\phi_i$, the hedging portfolio value $HV$ viewed as a function of the strike \emph{values} (with the number of strikes per asset and their sign pattern fixed) is continuous and piecewise-linear: the cash component $\overline{\lambda}$ is the minimum of a piecewise-linear function of $(x, \{K_i^j\})$ over a finite vertex grid, and each $\mathbb{E}[f_i(X_i)]$ is a linear combination of expected call payoffs that are themselves piecewise-linear in their strike on the discretised marginal. Consequently, a derivative-free local search initialised at the zero-crossings of the $\phi_i$ closes two sources of discrepancy: the finite-grid discretisation of the zero-crossing detection, and the misalignment between the objective used inside BRA and the actual $HV$ surface. We alternate the simplex method of \cite{nelder1965simplex} with the direction-set method of \cite{powell1964efficient} until neither improves $HV$. On this surface the simplex method alone stalls at kinks once there are more than a handful of free strikes (in the six-asset example of Section~\ref{S3} it stops up to $3\times10^{-5}$ short of the value reached by the alternation), and the line searches of Powell's method move along the kinks that stop it. The search may still end at a kink that is not a local maximum; this is harmless, because every strike configuration yields a valid sub-replicator and $HV$ never decreases along the search. The refinement therefore cannot overstate the bound: each candidate strike set yields a pointwise sub-replicator whose price is a valid lower bound for the given marginals, and the search merely selects the largest of these valid lower bounds. Note that $HV$ depends on the dependence structure only through the starting point of the search, since both $\mathbb{E}[f_i(X_i)]$ and $\overline{\lambda}$ are determined by the strikes and the marginals; the dependence structure enters through $BV$, the target of the certificate. In our numerical experiments this strike-refinement step typically closes the sub-replication gap to the reported precision for the dependence structure returned by BRA; see Section~\ref{S3}.

\begin{remark}[Alternative objectives inside RA/BRA]\label{rem:objectives}
Although we use the variance of the row-sum as the default objective for RA and BRA (as in \cite{bernard2018rearrangement}), each rearrangement step replaces the coupling of two block sums $U$ and $V$ by their antimonotone coupling, which minimises $\mathbb{E}[\varphi(U+V)]$ over all couplings of $U$ and $V$ for \emph{every} convex $\varphi$ \citep{tchen1980inequalities,ruschendorf1983solution}. Each step is therefore non-increasing in $\mathbb{E}[\varphi(\sum_i X_i)]$ for any convex $\varphi$, whatever objective is used to monitor convergence. In particular, taking $\varphi(z) = (z - K)^+$ yields a $K$-aware objective that directly targets the basket value $BV$ at the strike of interest. Running BRA with both the variance proxy and the $K$-aware objective and retaining the run that maximises $HV$ adds diversity to the set of candidate dependence structures at no additional theoretical cost.
\end{remark}

\begin{remark}[Portfolios without strikes]\label{rem:trivial}
Two sub-replicating portfolios require no strike search: the linear portfolio $\sum_i x_i-K$ of Proposition~\ref{prop:case1}, worth $m_X-K$ with $m_X:=\sum_i\mathbb{E}[X_i]$, and the portfolio $\sum_i(x_i-K)^+$ consisting of one call struck at $K$ on each asset, worth $\sum_iC_{X_i}(K)\ge0$ (it corresponds to $f_i(z)=(z-K)^+$, i.e.\ to the strikes $K_i^1=0$, $K_i^2=K$, with $\overline\lambda=0$). The implementation evaluates both before the strike search of each restart. If the better of the two already matches the basket value under the rearranged sample, it is optimal for the marginals used in the computation by Corollary~\ref{coll:optimality} and is returned at once; this is what happens when the basket can be kept on one side of the strike, in particular for jointly mixable marginals (Proposition~\ref{prop:case3}), a situation in which the functions $\phi_i$ carry no information because the basket is almost constant. Otherwise the strike search and the strike refinement proceed, and the refined portfolio is retained only if it is worth more than both. The comparison has to be made after the refinement and not before: at deep out-of-the-money strikes the value of the bound is of the order of the grid error of the zero-crossings, so that the unrefined strikes can be worth less than the strike-free portfolios although the refined ones attain the basket value (this is the case at $K=115$ and $K=120$ in the three-asset example of Section~\ref{S3}). The value returned by Algorithm~\ref{alg:subrep} is therefore never below $(m_X-K)^+$.
\end{remark}

\subsection{Bounds for the original marginals}\label{ssec:exact}

The rearrangement algorithms operate on marginals discretised into $N$ equiprobable atoms, so the equality $HV=BV$ of Corollary~\ref{coll:optimality} certifies optimality for the discretised problem (Remark~\ref{rem:certificate}). Two inexpensive additions to the construction transfer the certificate to the original problem \eqref{eq:primal}, one from each side.

\emph{Lower bound.} The portfolio $\Pi^{\mathrm{sub}}$ sub-replicates the basket payoff on all of $\mathbb{R}^n_+$ whatever its strikes, so its price under the original marginals,
$$
HV^{\mathrm{ex}}\bigl(\{K_i^j\}\bigr)\;=\;\sum_{i=1}^{n}\Bigl\{\mathbb{E}[X_i]+\sum_{j=1}^{2n_i}(-1)^{j}\,C_{X_i}\bigl(K_i^j\bigr)\Bigr\}+\overline{\lambda}\bigl(\{K_i^j\}\bigr),
$$
is a lower bound for \eqref{eq:primal}; here $C_{X_i}$ is the call price function of the original law and $\overline{\lambda}$ is the exact cash component of Proposition~\ref{prop:exact-lambda}. Instead of evaluating this expression at the strikes refined on the sample, we maximise it: the local search of Section~\ref{ssec:algorithm} is repeated with $HV^{\mathrm{ex}}$ as objective, started from the strikes obtained on the discretised marginals. The long calls that had been placed at $K$ because they were worthless on the sample (their strike exceeded the largest atom) have a positive price under the original law. The search is therefore run a second time with these calls released, starting from the largest atom, and the outcome of this second run is retained if it raises $HV^{\mathrm{ex}}$ by more than $10^{-6}$; below that threshold the value is practically flat in the strike of such a call, and the call is left at $K$. As before, every strike configuration visited is a valid sub-replicator, so the search can only improve a valid lower bound.

\emph{Upper bound.} The rearranged sample is a dependence structure of the discretised marginals, not of the original ones, so $BV$ is not an upper bound for \eqref{eq:primal}. It becomes one after each atom is replaced by the piece of the original law that it represents. Let $F_i^{-1}$ denote the quantile function of $X_i$, and let the rearranged sample consist of the rows $(x_1^{(\pi_1(m))},\ldots,x_n^{(\pi_n(m))})$, $m=1,\ldots,N$, where $x_i^{(1)}<\cdots<x_i^{(N)}$ are the atoms of asset $i$ and $\pi_1,\ldots,\pi_n$ are permutations of $\{1,\ldots,N\}$.

\begin{prop}[Reassembly with the original marginals]\label{prop:reassembly}
Let $M$ be uniformly distributed on $\{1,\ldots,N\}$ and $T$ uniformly distributed on $(0,1)$, independent of $M$. For every row $m$ and asset $i$ fix an orientation $\epsilon_{m,i}\in\{0,1\}$, put $T_i:=T$ if $\epsilon_{M,i}=0$ and $T_i:=1-T$ otherwise, and define
\begin{equation}\label{eq:reassembly}
\widetilde X_i\;:=\;F_i^{-1}\!\left(\frac{\pi_i(M)-1+T_i}{N}\right),\qquad i=1,\ldots,n .
\end{equation}
Then the law $\widetilde\mu$ of $(\widetilde X_1,\ldots,\widetilde X_n)$ belongs to $\Pi$, and consequently
\begin{equation}\label{eq:bvex}
BV^{\mathrm{ex}}\;:=\;\frac1N\sum_{m=1}^{N}\int_0^1\Bigl(\sum_{i=1}^{n}F_i^{-1}\Bigl(\tfrac{\pi_i(m)-1+t_i}{N}\Bigr)-K\Bigr)^{+}dt\;\;\ge\;\;\inf_{\mu\in\Pi}\mathbb{E}_\mu\Bigl[\Bigl(\sum_{i=1}^{n}X_i-K\Bigr)^{+}\Bigr],
\end{equation}
where $t_i=t$ if $\epsilon_{m,i}=0$ and $t_i=1-t$ otherwise.
\end{prop}

\begin{proof}
Fix $i$. Conditionally on $M=m$, the variable $T_i$ is uniform on $(0,1)$ whatever the orientation, so $U_i:=(\pi_i(M)-1+T_i)/N$ is uniform on the interval $\bigl((\pi_i(m)-1)/N,\,\pi_i(m)/N\bigr)$. Since $\pi_i$ is a permutation and $M$ is uniform, these $N$ intervals are visited with probability $1/N$ each and partition $(0,1)$ up to a null set; hence $U_i$ is uniform on $(0,1)$ and $\widetilde X_i=F_i^{-1}(U_i)$ has the law of $X_i$. Thus $\widetilde\mu\in\Pi$, and \eqref{eq:bvex} is the basket value under $\widetilde\mu$ written as an average over the rows.
\end{proof}

The construction is the one-dimensional case of the reassembly of \cite{neufeld2022mmot}, applied to the output of the rearrangement algorithm: each atom is spread over the slab of the original law lying between the quantile levels $(\pi_i(m)-1)/N$ and $\pi_i(m)/N$, and within a row the $n$ coordinates move along a curve driven by a single uniform variable. Two features make it convenient here. First, the orientations are free, so we select for each row the pattern in $\{0,1\}^n$ that gives the smallest row integral (the pattern and its mirror image give the same integral, which leaves $2^{n-1}$ candidates per row). Second, each row integral in \eqref{eq:bvex} is one-dimensional, so $BV^{\mathrm{ex}}$ is obtained by quadrature and not by simulation: we use a Gauss--Legendre rule on each row, with more nodes on the few rows along which the basket crosses the strike, and adaptive quadrature on the at most $2n$ rows that contain the lowest or the highest slab of some asset, where the quantile function or its derivative is unbounded.

\begin{coll}[Bracket for the original problem]\label{coll:bracket}
For any strikes $\{K_i^j\}$, any rearranged sample and any orientations,
$$
HV^{\mathrm{ex}}\;\le\;\inf_{\mu\in\Pi}\mathbb{E}_\mu\Bigl[\Bigl(\sum_{i=1}^{n}X_i-K\Bigr)^{+}\Bigr]\;\le\;BV^{\mathrm{ex}} .
$$
In particular the portfolio $\Pi^{\mathrm{sub}}$ and the dependence structure $\widetilde\mu$ are both optimal for the original marginals up to the computable gap $BV^{\mathrm{ex}}-HV^{\mathrm{ex}}$.
\end{coll}

\begin{proof}
The first inequality is weak duality for the pointwise sub-replicator $\Pi^{\mathrm{sub}}$ (Theorem~\ref{thm:feasible}), the second is Proposition~\ref{prop:reassembly}.
\end{proof}

Unlike Corollary~\ref{coll:optimality}, the bracket does not require an exact equality: the gap $BV^{\mathrm{ex}}-HV^{\mathrm{ex}}$ is a certified bound on the distance of both bounds from the true value of \eqref{eq:primal}, and it accounts for all three sources of error at once, namely the discretisation of the marginals, the possible sub-optimality of the BRA output and the possible sub-optimality of the strikes. Here $BV^{\mathrm{ex}}$
denotes the exact integral in Proposition \ref{prop:reassembly}.

\subsection{Feasible sub-replicating portfolio for basket put options}
	For completeness, we record the corresponding portfolio for basket put options. Since $(K-x_1-x_2-x_3)^+=(x_1+x_2+x_3-K)^+-(x_1+x_2+x_3)+K$ (put--call parity at the level of payoffs), subtracting $x_1+x_2+x_3-K$ from both sides of \eqref{eq:subrep} gives the following sub-replicating portfolio for the basket put:
	
$$
\tilde{f_{1}}\left( x_1\right) + \tilde{f_{2}}\left( x_2\right)+\tilde{f_{3}}\left( x_3\right) +\tilde{\lambda }
$$

\noindent where

 $$
\tilde{ f_{i}}\left( z\right) =\sum ^{n_i}_{a=1}\left\{ \left( z-K_i^{2a}\right) ^{+}-\left( z-K_i^{2a-1}\right) ^{+}\right\}
 $$

$$
\tilde{\lambda }=\overline{\lambda }+K
$$

\section{Closed-form lower bounds and optimal sub-replicating portfolios in special cases\label{S2}}

In this section we present three special cases in which the lower bound is available in closed form together with an optimal (rather than merely feasible) sub-replicating portfolio of vanilla options. Proofs are deferred to Appendix~\ref{App A}.

\subsection{Case 1: no active strikes}\label{ssec:case1}

\begin{prop}\label{prop:case1}
Suppose $A_i=\emptyset$ for at least one $i\in\{1,2,3\}$, i.e.\ $\phi_i\le 0$ on $\mathbb{R}_+$ for some $i$. (This holds in particular when $n_1=n_2=n_3=0$, in which case the construction of Section~\ref{S1} returns the portfolio below.) The sets $A_i$ may be computed from any dependence structure in $\Pi$; the conclusion does not depend on which one is used. Neither the statement nor its proof requires the marginals to have densities: both apply verbatim to the discretised marginals of the implementation, with $\Pi$ the set of their dependence structures, and this is how the proposition is used in Algorithm~\ref{alg:subrep}. Then the portfolio
$$
\Pi^{\mathrm{sub}}_1(x_1,x_2,x_3):=x_1+x_2+x_3-K
$$
is a pointwise sub-replicator of the basket call payoff, and its expected cost equals the lower bound $\inf_{\mu\in\Pi}\mathbb{E}_\mu[(X_1+X_2+X_3-K)^+]$, hence it is optimal.
\end{prop}

Write $m_X:=\mathbb{E}[X_1]+\mathbb{E}[X_2]+\mathbb{E}[X_3]$ for the common basket mean (common because every $\mu\in\Pi$ preserves the marginals). This case applies whenever the strike $K$ is small enough that the basket option is sufficiently in-the-money for the lower bound to coincide with $m_X-K$. Although elementary, the regime is common and merits separate treatment.

\subsection{Case 2: one dominant asset}\label{ssec:case2}

The second regime is the one in which a single asset, say $X_3$, is spread widely enough to dominate the remaining pair. Let $\mu^{\mathrm{c}}\in\Pi$ denote the dependence structure under which $(X_1,X_2)$ is co-monotone and $X_3$ is antimonotone to $X_1+X_2$. Under $\mu^{\mathrm{c}}$ the vector $(X_1,X_2,X_3)$ is supported on the curve $u\mapsto\bigl(F_{X_1}^{-1}(u),F_{X_2}^{-1}(u),F_{X_3}^{-1}(1-u)\bigr)$, $u\in(0,1)$, along which the basket equals
\begin{equation}\label{eq:Sc}
S^{\mathrm{c}}(u):=F_{X_1}^{-1}(u)+F_{X_2}^{-1}(u)+F_{X_3}^{-1}(1-u),
\end{equation}
a continuous function of $u$ when the distribution functions are continuous and strictly increasing, and $V(\mu^{\mathrm{c}}):=\mathbb{E}_{\mu^{\mathrm{c}}}[(X_1+X_2+X_3-K)^+]=\int_0^1(S^{\mathrm{c}}(u)-K)^+du$. Near $u=0$ the dominant asset is large and $S^{\mathrm{c}}$ exceeds $K$; the regime of interest is the one in which $S^{\mathrm{c}}$ crosses the level $K$ only once, from above. For $a_1,a_2,c\ge 0$ with $a_1+a_2+c=K$ define the vanilla portfolio
\begin{equation}\label{eq:pi2}
\Pi^{\mathrm{sub}}_2(x_1,x_2,x_3):=\min(x_1,a_1)+\min(x_2,a_2)+(x_3-c)^+-a_1-a_2,
\end{equation}
i.e.\ long positions in $X_1$ and $X_2$ capped by short calls struck at $a_1$ and $a_2$, a long call on $X_3$ struck at $c$, and a short cash position $a_1+a_2$. This is the portfolio of Section~\ref{S1} with $f_1(z)=\min(z,a_1)$, $f_2(z)=\min(z,a_2)$ and $f_3(z)=(z-c)^+$, that is, with strikes $K_1^1=a_1$, $K_2^1=a_2$, $K_3^1=0$ and $K_3^2=c$, the long calls $K_1^2$ and $K_2^2$ being struck at or above $K$ (where they may be added without affecting the argument below), and with $\overline{\lambda}=-(a_1+a_2)$. We write $P_{X_i}(k):=\mathbb{E}[(k-X_i)^+]$ for the put price function.

\begin{prop}\label{prop:case2}
Let the marginals have finite means and distribution functions that are continuous and strictly increasing on their supports (Assumption~\ref{ass:marginals} with the full-support part waived).
\begin{enumerate}[label=(\roman*)]
\item For every $a_1,a_2,c\ge 0$ with $a_1+a_2+c=K$, the portfolio \eqref{eq:pi2} is a pointwise sub-replicator of the basket call payoff on $\mathbb{R}^3_+$, the constant $-(a_1+a_2)$ is its exact cash component \eqref{eq:lambda-def}, and its expected cost under every $\mu\in\Pi$ is $C_{X_3}(c)-P_{X_1}(a_1)-P_{X_2}(a_2)$. Consequently
\begin{equation}\label{eq:case2-lb}
\inf_{\mu\in\Pi}\mathbb{E}_\mu\bigl[(X_1+X_2+X_3-K)^+\bigr]\;\ge\;\max_{a_1+a_2+c=K}\bigl\{C_{X_3}(c)-P_{X_1}(a_1)-P_{X_2}(a_2)\bigr\}.
\end{equation}
\item If $u^*\in(0,1)$ satisfies $S^{\mathrm{c}}(u^*)=K$, then $a_i:=F_{X_i}^{-1}(u^*)$ for $i=1,2$ and $c:=F_{X_3}^{-1}(1-u^*)$ satisfy $a_1+a_2+c=K$ and
$$
C_{X_3}(c)-P_{X_1}(a_1)-P_{X_2}(a_2)=\int_0^{u^*}\bigl(S^{\mathrm{c}}(u)-K\bigr)\,du .
$$
\item If moreover $S^{\mathrm{c}}\ge K$ on $(0,u^*]$ and $S^{\mathrm{c}}\le K$ on $[u^*,1)$, then $\mu^{\mathrm{c}}$ is optimal, the lower bound equals
$$
\inf_{\mu\in\Pi}\mathbb{E}_\mu\bigl[(X_1+X_2+X_3-K)^+\bigr]=V(\mu^{\mathrm{c}})=\int_0^{u^*}\bigl(S^{\mathrm{c}}(u)-K\bigr)\,du ,
$$
which equals $C_{X_3}(c)-P_{X_1}(a_1)-P_{X_2}(a_2)$ by (ii), and the portfolio \eqref{eq:pi2} with the strikes of (ii) is an optimal sub-replicating portfolio of vanilla options, with $\Pi^{\mathrm{sub}}_2=(X_1+X_2+X_3-K)^+$ $\mu^{\mathrm{c}}$-almost surely.
\end{enumerate}
\end{prop}

The single-crossing condition of (iii) is a condition on the marginals alone. Since $S^{\mathrm{c}}(1-)=\operatorname{ess\,sup}X_1+\operatorname{ess\,sup}X_2+\operatorname{ess\,inf}X_3$, it requires the two dominated assets to be bounded, with $\operatorname{ess\,sup}X_1+\operatorname{ess\,sup}X_2+\operatorname{ess\,inf}X_3\le K$; for the discretised marginals used in the implementation, which are bounded, it can be read off the $N$ values of $S^{\mathrm{c}}$ directly, and for unbounded marginals such as the log-normal it holds up to the mass of the region near $u=1$ where $S^{\mathrm{c}}$ rises above $K$ again. When $S^{\mathrm{c}}$ crosses $K$ more than once, $\mu^{\mathrm{c}}$ is in general no longer optimal, but the bound \eqref{eq:case2-lb} may still be attained: the portfolio \eqref{eq:pi2} coincides with the basket payoff on the two regions $\{x_1\le a_1,\,x_2\le a_2,\,x_3\ge c,\,\sum_i x_i\ge K\}$ and $\{x_1\ge a_1,\,x_2\ge a_2,\,x_3\le c,\,\sum_i x_i\le K\}$, so equality holds in \eqref{eq:case2-lb} as soon as some dependence structure in $\Pi$ is supported on their union, which Corollary~\ref{coll:optimality} tests numerically. Statements (i)--(iii) extend verbatim to $n$ assets, with $X_1,\ldots,X_{n-1}$ capped and a single call on the dominant asset $X_n$. The three situations are illustrated numerically in Section~\ref{ssec:case2-numerics}.

\begin{remark}[Relation to the two-asset problem]\label{rem:case2-hlw}
If $(X_1,X_2)$ is co-monotone under some optimal dependence structure, then $\mu^{\mathrm{c}}$ is optimal and the lower bound equals $V(\mu^{\mathrm{c}})$, because for the marginals of $X_1+X_2$ and $X_3$ the antimonotone coupling minimises $\mathbb{E}[(X_1+X_2+X_3-K)^+]$ \citep{tchen1980inequalities}; the three-asset problem then reduces to the two-asset problem of \cite{hobson2005sub} on the pair $(X_1+X_2,X_3)$. That reduction gives the value of the bound but not a hedge in the vanilla market: the two-asset portfolio contains calls on the sub-basket $X_1+X_2$, and replacing each of them by the vanilla combination $(x_1-a)^++(x_2-b)^+$ with $a+b$ equal to its strike, which has the same price under $\mu^{\mathrm{c}}$, super-replicates the sub-basket call and destroys pointwise sub-replication. Proposition~\ref{prop:case2}(iii) replaces the unverifiable hypothesis of co-monotonicity at the optimum by a condition on the marginals and delivers an optimal portfolio of vanilla options.
\end{remark}

\subsection{Case 3: jointly mixable marginals}\label{ssec:case3}

\begin{prop}\label{prop:case3}
Suppose the marginals are jointly mixable \citep{wang2016joint}, in the sense that there exists a dependence structure $\mu^*\in\Pi$ under which $X_1+X_2+X_3$ is $\mu^*$-almost surely constant. By marginal-mean preservation this constant is necessarily $m_X$. (Joint mixability of non-negative variables forces each $X_i$ to be bounded by $m_X$, so the full-support part of Assumption~\ref{ass:marginals} is waived in this case.) Then the lower bound of the basket call price is
$$
\inf_{\mu\in\Pi}\mathbb{E}_\mu\bigl[(X_1+X_2+X_3-K)^+\bigr]
=\bigl(m_X-K\bigr)^+.
$$
The optimal sub-replicating portfolio is:
\begin{itemize}
\item if $K\le m_X$, the portfolio $\Pi^{\mathrm{sub}}_1$ of Proposition~\ref{prop:case1} (i.e.\ $x_1+x_2+x_3-K$, with expected cost $m_X-K$);
\item if $K>m_X$, the trivial zero portfolio.
\end{itemize}
\end{prop}

\section{On the optimality of the construction\label{sec:optimality}}

The construction of Section~\ref{S1} is unconditionally feasible (Theorem~\ref{thm:feasible}), and Corollary~\ref{coll:optimality} provides a verifiable sufficient condition for optimality whenever the hedging-portfolio value coincides with the basket value under the BRA dependence structure, for the marginals used in the computation. The three special regimes of Section~\ref{S2} give the value of the lower bound in closed form. A natural follow-up question is whether this coincidence, and hence the optimality of the construction, can be guaranteed a priori for general marginals in dimension $n\ge 3$. The short answer is negative, and two distinct obstructions explain why.

The first obstruction concerns the origin of the strikes. The dual of \eqref{eq:primal} over integrable functions of the individual assets has, by the theorem of \cite{kellerer1984duality}, value equal to the lower bound; the portfolio of Theorem~\ref{thm:feasible} belongs to the subclass of static positions in vanilla calls on $X_1,\ldots,X_n$ together with cash. Its strikes, however, are inherited from $n$ separate two-asset problems: for each pair $(X_i,Y_i)$ the Hobson--Laurence--Wang triplet $(\lambda_i,\overline{\nu}_i,\overline{\nu}_{y_i})$ is optimal in a market in which calls on the aggregate $Y_i$ are traded, and the construction keeps the $X_i$-leg $\overline{\nu}_i$ of each triplet, while the role of the discarded $Y_i$-legs is taken over by the other $f_j$ and by the single constant $\overline{\lambda}$. Nothing guarantees that strikes obtained in this way, even after the strike-refinement step of Section~\ref{ssec:algorithm}, coincide with a dual optimiser of the $n$-asset problem: the dual optimum may require a different number of strikes per asset, or may not be attained by finitely many strikes at all. In dimension two this issue does not arise, because the pair decomposition is the problem itself.

The second obstruction is that, even setting the first aside, the worst-case dependence structure $\mu^*$ attaining $\inf_{\mu\in\Pi}\mathbb{E}_\mu[(\sum_i X_i - K)^+]$ has no closed-form description in dimension three or above (Section~\ref{ssec:algorithm}), and characterising it remains an open problem in dependence modelling. BRA provides a numerical approximation but, as a heuristic, comes with no a priori guarantee of converging to a global minimiser; multiple restarts, the $K$-aware objective of Remark~\ref{rem:objectives} and the $HV$-maximising selection rule of Section~\ref{ssec:algorithm} mitigate this in practice, yet cannot deliver a theoretical guarantee.

The three special regimes of Section~\ref{S2} are precisely those in which one or both obstructions disappear. In Case~1 ($A_i=\emptyset$ for some $i$) the lower bound is attained by a dependence structure under which the basket payoff is linear, so the linear portfolio $x_1+x_2+x_3-K$ is dual optimal and BRA uncertainty plays no role. In Case~2 (one dominant asset) the basket along the co-monotone/antimonotone coupling $\mu^{\mathrm{c}}$ crosses the strike once, from above; the worst-case dependence structure is then $\mu^{\mathrm{c}}$ itself, and the vanilla portfolio \eqref{eq:pi2}, which has the pair-decomposition form with a single strike on each dominated asset and a single call on the dominant one, is dual optimal, so both obstructions disappear. In Case~3 (joint mixability) the basket collapses to a deterministic constant, trivialising the lower bound. Outside these three regimes neither obstruction is known to vanish, and we are not aware of a general analytical optimality result for $n\ge 3$ baskets in the literature.

The two obstructions are theoretical and, to the best of our knowledge, cannot be removed without solving the open problem of characterising basket-option lower bounds in dimensions $\ge 3$. Corollary~\ref{coll:optimality} provides a complementary form of progress by reducing the question of optimality to a single numerical comparison between the hedging-portfolio value and the basket value under the BRA dependence structure, with both quantities computed exactly (free of Monte-Carlo noise) once Proposition~\ref{prop:exact-lambda} is invoked, for the discretised marginals used in the computation. Whenever this equality is observed, as it is to the reported precision at every strike of the three-asset example of Section~\ref{S3}, weak duality shows that at that strike the pair-decomposition strikes are dual optimal and the BRA output is primal optimal for the discretised problem, even though such an outcome cannot be anticipated analytically. Corollary~\ref{coll:bracket} extends this to the original marginals and dispenses with the equality: the difference between $BV^{\mathrm{ex}}$ and $HV^{\mathrm{ex}}$ bounds the sub-optimality of the portfolio and of the reassembled dependence structure, whatever its origin. The contribution of the construction is therefore best described as verifiably optimal rather than provably optimal: it returns a feasible sub-replicator, a dependence structure of the original marginals, and a certified bound on the distance of each from the optimum, which is at most $0.1\%$ of the value of the bound in the examples of Section~\ref{S3}.

\section{Numerical examples \label{S3}}
In this section we illustrate the algorithm in the Black--Scholes setting.\footnote{A Python implementation of the algorithm, together with the scripts that reproduce the numerical examples of this section, is available at \url{https://github.com/MohamadAbouDaya/Subreplication}.} We assume that stock prices follow
$$
S_{i,T}= S_{i,0}\exp\!\Big(\sigma_i W_{i,T}-\tfrac{1}{2}\sigma_i^2 T\Big),
$$
where $\sigma_i$ is the volatility of asset $i$ and $W_i$ is a standard Brownian motion. As before, we use BRA to determine the dependence structure of the asset prices at time $T$, so we do not pre-specify any correlation structure between the $W_i$. We consider two examples with three and six assets respectively, and in Section~\ref{ssec:case2-numerics} three further three-asset examples illustrating Case~2. In both examples $S_{i,0}=100$ for every asset, the assets are equally weighted ($w_i=1/n$, so that the basket forward is $100$), and $T=0.5$. Table~\ref{tab:vol} reports the volatility of each asset. We discretise each marginal distribution into $50{,}000$ equiprobable atoms placed at the mid-point quantiles of the log-normal law before applying BRA; the cash component is computed exactly via Proposition~\ref{prop:exact-lambda}. The bounds for the original log-normal marginals are then computed as described in Section~\ref{ssec:exact}.

Tables~\ref{tab:example1} and~\ref{tab:example2} report, for each basket strike $K$, the sub-replicating portfolio returned by Algorithm~\ref{alg:subrep} and four values. The strike and cash columns describe the final portfolio, that is, the strikes obtained from the zero-crossings of the corresponding $\phi_i$ of \eqref{phi1}--\eqref{phi3}, refined first on the discretised marginals and then against the exact log-normal call prices (Section~\ref{ssec:exact}), together with the exact cash component of Proposition~\ref{prop:exact-lambda}. These strikes refer to the weighted asset $x_i = w_i S_i$. For example, at $K=100$ the construction calls for a long position in $x_1$, a short call on $x_1$ struck at $13.4746$, and a long call on $x_1$ struck at $50.1780$; equivalently, a $w_1$ position in $S_1$, a short $w_1$ position in a call on $S_1$ struck at $40.424$, and a long $w_1$ position in a call on $S_1$ struck at $150.534$. A strike equal to $K$ indicates a long call that is struck above the largest atom of the asset, hence worthless on the discretised marginal, and that the refinement against the exact prices left at $K$ because releasing it would raise the bound by less than $10^{-6}$ (Section~\ref{ssec:exact}). The strikes are rounded to four decimal places; the values are computed from the unrounded strikes, which are available with the code. The distinction matters: several vertices are tied in the minimum that defines the cash component of an optimal portfolio, so that rebuilding the portfolio from the rounded strikes can lower its value by up to $10^{-4}$. The columns $HV$ and $BV$ are the hedging value after the first refinement and the basket value under the selected BRA dependence structure, both on the discretised marginals; their equality certifies optimality for the discretised problem (Corollary~\ref{coll:optimality} and Remark~\ref{rem:certificate}). The last two columns concern the original problem \eqref{eq:primal}: $HV^{\mathrm{ex}}$ is the price of the tabulated portfolio under the exact log-normal marginals (Black formula), and $BV^{\mathrm{ex}}$ is the basket value of the BRA dependence structure reassembled with the exact marginals (Proposition~\ref{prop:reassembly}), so that the lower bound of the basket call in the log-normal model lies between them (Corollary~\ref{coll:bracket}). For $K$ below the boundary reported in the first row of each table the optimal sub-replicating portfolio consists of an equally-weighted long position in all underlying assets and a short cash position equal to $K$ (Case~1 of Section~\ref{S2}).

On the discretised marginals, $HV$ and $BV$ agree to five decimal places at every strike of the three-asset example, which activates the conclusion of Corollary~\ref{coll:optimality}: the BRA dependence structure attains the infimum of the discretised problem and the portfolio refined on the sample is optimal for it. In the six-asset example the two differ by $1\times10^{-6}$ to $2\times10^{-5}$. For the original marginals, the width $BV^{\mathrm{ex}}-HV^{\mathrm{ex}}$ of the bracket is between $2\times10^{-8}$ and $2\times10^{-5}$ in the three-asset example and between $1\times10^{-6}$ and $4\times10^{-5}$ in the six-asset example, that is, at most $0.1\%$ and $0.02\%$ of the value of the bound respectively. The lower bound of the basket call in the log-normal model is therefore determined to that accuracy, and the tabulated portfolios and the reassembled dependence structures are optimal up to the same margin. The discretisation of the marginals alone moves the values by more than this: $HV^{\mathrm{ex}}-HV$ is between $3\times10^{-5}$ and $5\times10^{-5}$ in the three-asset example and about $5\times10^{-4}$ in the six-asset example, where three assets have volatilities above $75\%$ and the truncation of their upper tails by the discretisation is more severe; at deep out-of-the-money strikes this is a non-negligible fraction of the option value itself (about $11\%$ at $K=120$ in the three-asset example). The bracket $[HV^{\mathrm{ex}},BV^{\mathrm{ex}}]$, not the pair $(HV,BV)$, is therefore the statement to quote for the log-normal model. The refinement against the exact prices adds little to the refinement on the sample. In the three-asset example it matters at two strikes: at $K=100$ it moves the long call on asset~2 from $K$ to $62.46$ and raises the lower bound from $0.102174$ to $0.102180$, and at $K=120$ its gain of $7\times10^{-7}$ closes most of what was left of the bracket; at the other strikes it changes the bound by less than $10^{-7}$. In the six-asset example it raises the bound by $1\times10^{-5}$ at $K=120$ and by $8\times10^{-6}$ at $K=130$, and by less than $3\times10^{-6}$ elsewhere. Recomputing $BV^{\mathrm{ex}}$ with four times as many quadrature nodes changes it by less than $10^{-10}$.

\begin{table}[hbt]
	\centering
	\caption{Asset volatility}
	\label{tab:vol}
	\begin{tabular}{ccc}
		\hline
		& \multicolumn{2}{c}{$\sigma$ (\%)} \\
		Asset & Example 1   & Example 2   \\ \hline\hline
		1     & 35.5        & 91.5        \\
		2     & 20          & 75.5        \\
		3     & 25          & 82.5        \\
		4     &             & 35.5        \\
		5     &             & 20          \\
		6     &             & 15          \\ \hline
	\end{tabular}
\end{table}

	\begin{table}[]
		\centering
		\footnotesize
		\caption{Sub-replicating portfolio and bounds for the basket call option on three log-normal assets. Strikes and cash describe the final portfolio of Algorithm~\ref{alg:subrep}; $HV$ and $BV$ are computed on the discretised marginals ($N=50{,}000$), $HV^{\mathrm{ex}}$ and $BV^{\mathrm{ex}}$ under the exact log-normal marginals.}
		\label{tab:example1}
		\setlength{\tabcolsep}{3pt}
		\begin{tabular}{ccccccccc}
			\hline
			\multirow{2}{*}{$K$} & \multicolumn{3}{c}{Asset Strikes}                 & \multirow{2}{*}{Cash} & \multirow{2}{*}{$HV^{*}$} & \multirow{2}{*}{$BV^{**}$} & \multirow{2}{*}{$HV^{\mathrm{ex}}$} & \multirow{2}{*}{$BV^{\mathrm{ex}}$} \\
			& 1                & 2            & 3               &                       &                                            &                               & & \\ \hline\hline
			\textless{}99.89               & \multicolumn{8}{c}{CASE 1}                                                                                                                             \\
			100                            & 13.4746; 50.1780 & 25.7590; 62.4624 & 24.0630; 60.7664 & -63.2966 & 0.10213 & 0.10213 & 0.10218 & 0.10220 \\
			105                            & 0; 60.9324 & 23.0803; 105 & 20.9873; 105 & -44.0676 & 0.01807 & 0.01807 & 0.01810 & 0.01811 \\
			110                            & 0; 69.4024 & 21.4482; 110 & 19.1494; 110 & -40.5976 & 0.00401 & 0.00401 & 0.00404 & 0.00404 \\
			115                            & 0; 76.9346 & 20.2555; 115 & 17.8099; 115 & -38.0654 & 0.00099 & 0.00099 & 0.00102 & 0.00102 \\
			120                            & 0; 83.9968 & 19.2617; 120 & 16.7414; 120 & -36.0032 & 0.00025 & 0.00025 & 0.00028 & 0.00028 \\ \hline
		\end{tabular}
		\begin{tablenotes}
			\item * Hedging portfolio value on the discretised marginals (after the first strike refinement)
			\item ** Basket value under the selected BRA dependence structure (discretised marginals)
			\item $HV^{\mathrm{ex}}$: price of the tabulated portfolio under the exact log-normal marginals (Black formula); a lower bound for the original problem \eqref{eq:primal}
			\item $BV^{\mathrm{ex}}$: basket value of the BRA dependence structure reassembled with the exact marginals (Proposition~\ref{prop:reassembly}); an upper bound for the original problem \eqref{eq:primal}
		\end{tablenotes}

	\end{table}

\begin{table}[]
	\centering
	\footnotesize
	\caption{Sub-replicating portfolio and bounds for the basket call option on six log-normal assets. Strikes and cash describe the final portfolio of Algorithm~\ref{alg:subrep}; $HV$ and $BV$ are computed on the discretised marginals ($N=50{,}000$), $HV^{\mathrm{ex}}$ and $BV^{\mathrm{ex}}$ under the exact log-normal marginals (notes as in Table~\ref{tab:example1}).}
	\label{tab:example2}
	\setlength{\tabcolsep}{2.5pt}
	\begin{tabular}{cccccccccccc}
		\hline
		\multirow{2}{*}{$K$} & \multicolumn{6}{c}{Asset Strikes} & \multirow{2}{*}{Cash} & \multirow{2}{*}{$HV$} & \multirow{2}{*}{$BV$} & \multirow{2}{*}{$HV^{\mathrm{ex}}$} & \multirow{2}{*}{$BV^{\mathrm{ex}}$} \\
		& 1 & 2 & 3 & 4 & 5 & 6 & & & & & \\ \hline\hline
		\textless{}99.66               & \multicolumn{11}{c}{CASE 1} \\
		100 & \begin{tabular}[c]{@{}c@{}}3.3300;\\ 54.2087\end{tabular} & \begin{tabular}[c]{@{}c@{}}5.0959;\\ 55.9746\end{tabular} & \begin{tabular}[c]{@{}c@{}}4.3135;\\ 55.1922\end{tabular} & \begin{tabular}[c]{@{}c@{}}10.1079;\\ 100\end{tabular} & \begin{tabular}[c]{@{}c@{}}12.6764;\\ 100\end{tabular} & \begin{tabular}[c]{@{}c@{}}13.5975;\\ 100\end{tabular} & -49.1213 & 0.32679 & 0.32679 & 0.32733 & 0.32737 \\
		110 & \begin{tabular}[c]{@{}c@{}}2.5117;\\ 68.4893\end{tabular} & \begin{tabular}[c]{@{}c@{}}4.1050;\\ 70.0826\end{tabular} & \begin{tabular}[c]{@{}c@{}}3.4252;\\ 69.4028\end{tabular} & \begin{tabular}[c]{@{}c@{}}9.0722;\\ 110\end{tabular} & \begin{tabular}[c]{@{}c@{}}11.9232;\\ 110\end{tabular} & \begin{tabular}[c]{@{}c@{}}12.9851;\\ 110\end{tabular} & -44.0224 & 0.14060 & 0.14060 & 0.14114 & 0.14114 \\
		120 & \begin{tabular}[c]{@{}c@{}}2.0564;\\ 81.1217\end{tabular} & \begin{tabular}[c]{@{}c@{}}3.5228;\\ 82.5880\end{tabular} & \begin{tabular}[c]{@{}c@{}}2.9287;\\ 81.9939\end{tabular} & \begin{tabular}[c]{@{}c@{}}8.4218;\\ 120\end{tabular} & \begin{tabular}[c]{@{}c@{}}11.4440;\\ 120\end{tabular} & \begin{tabular}[c]{@{}c@{}}12.5611;\\ 120\end{tabular} & -40.9348 & 0.06928 & 0.06930 & 0.06983 & 0.06984 \\
		130 & \begin{tabular}[c]{@{}c@{}}1.7405;\\ 93.1636\end{tabular} & \begin{tabular}[c]{@{}c@{}}3.1034;\\ 94.5264\end{tabular} & \begin{tabular}[c]{@{}c@{}}2.5328;\\ 93.9559\end{tabular} & \begin{tabular}[c]{@{}c@{}}7.8994;\\ 130\end{tabular} & \begin{tabular}[c]{@{}c@{}}11.0310;\\ 130\end{tabular} & \begin{tabular}[c]{@{}c@{}}12.2698;\\ 130\end{tabular} & -38.5770 & 0.03703 & 0.03704 & 0.03758 & 0.03758 \\
		140 & \begin{tabular}[c]{@{}c@{}}1.5068;\\ 104.6982\end{tabular} & \begin{tabular}[c]{@{}c@{}}2.7984;\\ 105.9899\end{tabular} & \begin{tabular}[c]{@{}c@{}}2.2703;\\ 105.4618\end{tabular} & \begin{tabular}[c]{@{}c@{}}7.5119;\\ 140\end{tabular} & \begin{tabular}[c]{@{}c@{}}10.7165;\\ 140\end{tabular} & \begin{tabular}[c]{@{}c@{}}12.0046;\\ 140\end{tabular} & -36.8085 & 0.02092 & 0.02092 & 0.02146 & 0.02146 \\ \hline
	\end{tabular}
\end{table}

\FloatBarrier
\subsection{One dominant asset (Case 2)}\label{ssec:case2-numerics}

Table~\ref{tab:case2} illustrates Proposition~\ref{prop:case2} in four three-asset examples, each marginal being discretised into $N=50{,}000$ mid-point quantiles as before; $BV$ denotes the smallest basket value found by RA and BRA. In the first two rows one asset dominates, $S^{\mathrm{c}}$ crosses $K$ once, from above, and the bound \eqref{eq:case2-lb} coincides with both $V(\mu^{\mathrm{c}})$ and $BV$, so that Corollary~\ref{coll:optimality} certifies the optimality of $\mu^{\mathrm{c}}$ and of the portfolio \eqref{eq:pi2}, as Proposition~\ref{prop:case2}(iii) predicts. In the third row $S^{\mathrm{c}}$ crosses $K$ twice, so that (iii) does not apply, and $\mu^{\mathrm{c}}$ is indeed not optimal, since $BV<V(\mu^{\mathrm{c}})$; the bound \eqref{eq:case2-lb} nevertheless coincides with $BV$, so that the portfolio \eqref{eq:pi2} built from the first downcrossing is optimal although $\mu^{\mathrm{c}}$ is not. In the last row, the three-asset example of Table~\ref{tab:example1}, no asset dominates, $S^{\mathrm{c}}$ crosses $K$ from below, and the regime does not apply. In the log-normal rows the exact marginals are unbounded, so the single-crossing condition of Proposition~\ref{prop:case2}(iii) holds for the discretised marginals and, for the exact ones, up to the mass of the region near $u=1$ where $S^{\mathrm{c}}$ rises above $K$ again; as for Tables~\ref{tab:example1}--\ref{tab:example2}, the certificate concerns the discretised problem.

\begin{table}[hbt]
\centering
\footnotesize
\caption{Case 2 in four three-asset examples on discretised marginals ($N=50{,}000$ mid-point quantiles). Log-normal rows use the setting of this section ($S_{i,0}=100$, $w_i=1/3$, $T=0.5$) with the volatilities shown; in the first row $X_3$ is log-normal with mean $40$ and volatility $100\%$ over $T=1$. $u^*$ is the first downcrossing of $S^{\mathrm{c}}$, ``bound'' is the right-hand side of \eqref{eq:case2-lb} at the strikes of Proposition~\ref{prop:case2}(ii), and $BV$ is the smallest basket value found by RA and BRA (both objectives of Remark~\ref{rem:objectives}, four restarts each).}
\label{tab:case2}
\setlength{\tabcolsep}{4pt}
\begin{tabular}{lcccccc}
\hline
Marginals & $K$ & Crossings of $S^{\mathrm{c}}$ & $u^*$ & $V(\mu^{\mathrm{c}})$ & Bound \eqref{eq:case2-lb} & $BV$ \\ \hline\hline
$\mathrm{U}[0,30]$, $\mathrm{U}[0,30]$, log-normal & 70  & one, from above & 0.1885 & 9.723356 & 9.723356 & 9.723356 \\
log-normal, $\sigma=(10\%,10\%,120\%)$   & 110 & one, from above & 0.2012 & 6.810342 & 6.810342 & 6.810342 \\
log-normal, $\sigma=(20\%,20\%,60\%)$    & 105 & two             & 0.1370 & 1.240672 & 1.223466 & 1.223466 \\
log-normal, $\sigma=(35.5\%,20\%,25\%)$  & 100 & one, from below & --     & 2.935512 & --       & 0.102129 \\ \hline
\end{tabular}
\end{table}

\subsection{Comparison with a general-purpose optimal-transport algorithm}\label{ssec:nx}

The multi-marginal optimal-transport algorithm of \cite{neufeld2022mmot} applies to problem \eqref{eq:primal}: the basket payoff is a Lipschitz, piecewise-affine cost and the marginals are one-dimensional, which is the setting of their third experiment. Their method replaces each marginal constraint by finitely many moment constraints with respect to the piecewise-linear ``hat'' functions on a grid of knots, solves the resulting semi-infinite linear programme by a cutting-plane scheme, and thereby obtains a lower bound together with a sub-replicating portfolio of vanilla options whose strikes are the knots; an upper bound is obtained by reassembling the discrete optimiser of the relaxed problem with the marginals and, in their implementation for this class of cost functions, is evaluated by Monte Carlo. We ran the authors' implementation on the two examples of Tables~\ref{tab:example1}--\ref{tab:example2}.\footnote{Code from \url{https://github.com/qikunxiang/MultiMarginalOptimalTransport}, class \texttt{MMOT1DCPWA}, with a log-normal marginal class added. The implementation calls the Gurobi solver, which we did not have; we replaced these calls by MATLAB's \texttt{linprog} and \texttt{intlinprog}, with HiGHS as a fallback whenever the linear-programming solution failed the optimality conditions, and left the algorithm itself unchanged. Every lower bound reported here was checked independently of that code: the dual solution was re-priced and the minimum of the basket payoff minus the portfolio over the support was recomputed by exact vertex enumeration (three assets) and by mixed-integer programming, the violation being at most $3\times10^{-7}$. The scripts are in the repository cited above. Running times obtained with the substitute solvers are not representative of the original implementation and are not reported.} Their algorithm and its error bounds are formulated for compactly supported marginals, so each log-normal law was truncated at its $1-10^{-12}$ quantile; the tolerance of the semi-infinite programme was $10^{-6}$, the knots were placed either at equally spaced quantile levels (``equal-probability'' knots, as in their third experiment) or at equally spaced points (as in their first experiment and in their bound on the number of test functions needed for a given accuracy; for this placement the marginals were truncated at the $10^{-8}$ and $1-10^{-8}$ quantiles, since knots in regions of negligible mass make the linear programmes ill-conditioned), and the upper bound was estimated from ten repetitions of $10^{6}$ samples.

\begin{table}[hbt]
\centering
\footnotesize
\caption{Bounds of \cite{neufeld2022mmot} with $m$ knots per asset, against the bracket of Tables~\ref{tab:example1}--\ref{tab:example2}. LB: lower bound (value of the sub-replicating portfolio with strikes at the knots); UB: upper bound (reassembly, Monte Carlo; standard errors at most $4\times10^{-4}$ for three assets and $1.5\times10^{-3}$ for six).}
\label{tab:nx}
\setlength{\tabcolsep}{5pt}
\begin{tabular}{ccccccc}
\hline
 & \multicolumn{4}{c}{\cite{neufeld2022mmot}} & \multicolumn{2}{c}{This paper} \\
 & \multicolumn{2}{c}{equal-probability knots} & \multicolumn{2}{c}{equally spaced knots} & & \\
$K$ & LB & UB & LB & UB & $HV^{\mathrm{ex}}$ & $BV^{\mathrm{ex}}$ \\ \hline\hline
\multicolumn{7}{l}{\emph{Three assets} ($m=256$ equal-probability knots, $m=64$ equally spaced knots)} \\
100 & 0.10163 & 0.12078 & 0.09849 & 0.30615 & 0.10218 & 0.10220 \\
105 & 0.01549 & 0.02557 & 0.01707 & 0.07611 & 0.01810 & 0.01811 \\
110 & 0       & 0.01705 & 0.00386 & 0.03034 & 0.00404 & 0.00404 \\
115 & 0       & 0.00382 & 0.00094 & 0.02085 & 0.00102 & 0.00102 \\
120 & 0       & 0.00633 & 0.00028 & 0.00572 & 0.00028 & 0.00028 \\ \hline
\multicolumn{7}{l}{\emph{Six assets} ($m=64$ equal-probability knots, $m=32$ equally spaced knots)} \\
100 & 0.20573 & 0.58989 & 0.19288 & 2.45440 & 0.32733 & 0.32737 \\
110 & 0       & 0.26336 & 0.09225 & 1.53732 & 0.14114 & 0.14114 \\
120 & 0       & 0.18060 & 0.04668 & 1.06603 & 0.06983 & 0.06984 \\
130 & 0       & 0.13617 & 0.02862 & 0.81591 & 0.03758 & 0.03758 \\
140 & 0       & 0.05381 & 0.01693 & 0.41734 & 0.02146 & 0.02146 \\ \hline
\end{tabular}
\end{table}

Table~\ref{tab:nx} shows the outcome. With equal-probability knots the lower bound of \cite{neufeld2022mmot} is weak out of the money: in the three-asset example with $256$ knots per asset, that is, a portfolio of up to $768$ calls, it is $0.10163$ at $K=100$, against $0.10218$ for our portfolio of six calls, and it is zero for $K\ge110$; in the six-asset example with $64$ knots per asset it is positive at $K=100$ only. The reason is the location of the knots. Out of the money the value of the bound is carried by the tails of the marginals, where equally spaced quantile levels place very few knots. Equally spaced knots cover the tails, and with $64$ of them per asset the lower bound is positive at every strike of the three-asset example and within $1\%$ to $8\%$ of ours ($0.000275$ against $0.000278$ at $K=120$, $0.00094$ against $0.00102$ at $K=115$); in the six-asset example with $32$ equally spaced knots per asset it is positive at every strike as well, and $21\%$ to $41\%$ below ours. With either placement the bound remains below the value of the portfolio built from the pair decomposition, which places one strike of each asset in the lower tail and one far in the upper tail. The upper bounds are less accurate than the lower ones: they exceed ours by $18\%$ at best ($K=100$, equal-probability knots) and by a factor of twenty at $K=120$, so that the certified interval for the value of the bound is wide in every case. With the substitute solvers the cutting-plane scheme converged slowly for equally spaced knots (several hundred iterations at $m=64$) and did not reach the tolerance within forty minutes with $m=128$ at $K=100$ in the three-asset example or with $m=64$ in the six-asset example; we therefore report $m=64$ and $m=32$ respectively (at $K=105$ in the three-asset example the run with $m=128$ did converge, to a lower bound of $0.01781$).

The role of the strike locations is confirmed by a second experiment, in which the strikes of Tables~\ref{tab:example1}--\ref{tab:example2} are added to a coarse grid of four equal-probability knots per asset. On a given grid the lower bound of \cite{neufeld2022mmot} is, up to the tolerance of their programme, the value of the best sub-replicating portfolio with strikes at the knots. This value is the optimum of a finite linear programme, because the sub-replication constraint needs to hold only at the vertices of the cells cut out by the knots and by the hyperplane $\{\sum_i x_i=K\}$, as in the proof of Proposition~\ref{prop:exact-lambda}; we solve it to optimality.\footnote{By constraint generation, with an exact enumeration of the vertices as separation step (script \texttt{augmented\_grid\_lp.py} in the repository). On the twenty grids on which we also ran the implementation of \cite{neufeld2022mmot}, with the strikes of a preliminary refinement, the optimum of the linear programme and the lower bound returned by that implementation differ by at most $3\times10^{-7}$, which is within its tolerance.} The best portfolio on the augmented grid then has the value $HV^{\mathrm{ex}}$ to within $3\times10^{-9}$, the effect of truncating the marginals, at every strike of both examples, and this does not change when the coarse grid is refined to sixteen knots. Read in one direction, this says that two well-placed strikes per asset achieve more than the grids of Table~\ref{tab:nx}. Read in the other, it is an independent check of our portfolios: among all sub-replicating portfolios with strikes on the augmented grid there is none better.

The two approaches are thus complementary. The algorithm of \cite{neufeld2022mmot} handles general cost functions, comes with bounds that provably tighten as the grid is refined, and is designed for many marginals (their third experiment has one hundred, and the complexity of their relaxed problem is polynomial in the number of marginals when the dimensions are fixed and the global minimisation over the product space is tractable), none of which holds for our construction. For the basket payoff, the construction of this paper supplies what a grid does not, namely the location of the strikes, and it returns a portfolio that sub-replicates on all of $\mathbb{R}^n_+$ without truncating the marginals, together with an upper bound computed by quadrature whose distance to the lower bound is smaller than in Table~\ref{tab:nx} by a factor of at least $600$, and by four orders of magnitude or more at most strikes.

\subsection{An example with the marginals of Neufeld and Xiang}\label{ssec:nx-marginals}

To compare the two methods on marginals that are not log-normal and on a problem that is not of our choosing, we take the marginals of the third experiment of \cite{neufeld2022mmot}, which are mixtures of three to five normal distributions truncated to $[-10,10]$, generated by their script, and one convex term of their cost function, $|\langle s,x\rangle-t|$. We restrict it to their first three marginals, with $s=(0.4739,-0.8802,-0.0259)$ the corresponding entries of their first direction vector rescaled to unit length, and $t=\mathbb{E}\langle s,X\rangle+\delta$ for $\delta\in\{0,1,2\}$, where $\mathbb{E}\langle s,X\rangle=-1.0557$. Since $|z-t|=2(z-t)^+-(z-t)$, bounds for this cost follow from bounds for a basket call on the non-negative assets $y_i=s_ix_i+10|s_i|$, to which Algorithm~\ref{alg:subrep} applies with the call prices and quantile functions of the truncated mixtures in place of the log-normal ones. Their method is run on its own marginal class with its own knot placement.

\begin{table}[hbt]
\centering
\footnotesize
\caption{Bounds on $\inf_{\mu\in\Pi}\mathbb{E}_\mu|\langle s,X\rangle-t|$ for three marginals of the third experiment of \cite{neufeld2022mmot}. Their bounds use $64$ equal-probability knots per marginal (upper bound by Monte Carlo, standard errors at most $6\times10^{-4}$).}
\label{tab:nx-marginals}
\setlength{\tabcolsep}{6pt}
\begin{tabular}{ccccc}
\hline
 & \multicolumn{2}{c}{This paper} & \multicolumn{2}{c}{\cite{neufeld2022mmot}} \\
$\delta$ & lower & upper & lower & upper \\ \hline\hline
0 & 1.693679 & 1.693679 & 1.693471 & 1.69517 \\
1 & 1.913807 & 1.913807 & 1.913589 & 1.91895 \\
2 & 2.355330 & 2.355330 & 2.355084 & 2.35821 \\ \hline
\end{tabular}
\end{table}

Table~\ref{tab:nx-marginals} reports the results. On these bounded, light-tailed marginals the grid-based lower bound is much closer to the truth than in the log-normal examples, within $3\times10^{-4}$ at $64$ knots per marginal, and their upper bound is within $6\times10^{-3}$. Our bracket closes: the two bounds differ by less than $10^{-11}$, which is the accuracy of the quadrature, so the BRA dependence structure and the portfolio are optimal for the original marginals to that accuracy. The example also shows where the basket structure matters. With six of these marginals and unit weights the problem becomes degenerate: the marginals are jointly mixable to numerical accuracy, the lower bound of the basket call is $(m_X-K)^+$, and the strike-free portfolios of Remark~\ref{rem:trivial} are returned, with brackets of width $3\times10^{-6}$ just below the mean ($K=5$, $m_X=5.118$) and below $2\times10^{-8}$ above it.

\subsection{Performance and limitations}

The quality of the output is measured by the width of the bracket of Corollary~\ref{coll:bracket} relative to the value of the bound,
\begin{equation}\label{eq:gap}
\varepsilon := \frac{BV^{\mathrm{ex}} - HV^{\mathrm{ex}}}{BV^{\mathrm{ex}}},
\end{equation}
which is at most $0.1\%$ in the three-asset example and $0.02\%$ in the six-asset example. Three ingredients contribute to this: the exact computation of $\overline{\lambda}$ via Proposition~\ref{prop:exact-lambda}, which removes the Monte-Carlo error of the cash component; the selection of the BRA output by its hedging value rather than by the variance of the sum, together with the strike refinement; and the transfer of both bounds to the original marginals in Section~\ref{ssec:exact}. On a sixteen-core machine a strike of the three-asset example takes about twenty seconds with eight restarts per configuration and a strike of the six-asset example about two and a half minutes with four restarts. The tables use $64$ and $20$ restarts respectively; with the smaller numbers of restarts the four reported values change by at most $3\times10^{-7}$ in the three-asset example and $8\times10^{-6}$ in the six-asset example, and the strikes by at most $0.06$, the value of the portfolio being nearly flat along some directions in the space of strikes.

Several challenges arise when extending the method to baskets with many assets. First, BRA is a heuristic that may converge to a local minimum (Section~\ref{ssec:algorithm}). Running it multiple times and selecting by $HV$ mitigates this risk but increases computation time; whatever the outcome, the bracket of Corollary~\ref{coll:bracket} reports how far the result is from the optimum.

Second, the computational cost of BRA grows with dimension. Each iteration operates on $N\times n$ entries (where $N$ is the number of atoms and $n$ the number of assets), and the number of restarts needed to avoid local minima may also increase with $n$.

Third, the cost of the exact computation of $\overline{\lambda}$ via Proposition~\ref{prop:exact-lambda} grows as $\prod_{i=1}^{n}(2n_i+2)$, which is exponential in $n$, and the reassembly of Proposition~\ref{prop:reassembly} examines $2^{n-1}$ orientation patterns per row. For the basket sizes considered in this paper ($n\le 6$, with $n_i=1$ in every reported row), this remains tractable ($4^6=4{,}096$ evaluations). For much larger baskets or assets with many active strikes, one could solve the piecewise-linear minimisation via mixed-integer programming, as \cite{neufeld2022mmot} do, and restrict the reassembly to a few orientation patterns, any of which gives a valid upper bound; in that regime their algorithm, which is designed for a large number of marginals, is the natural alternative.

Fourth, the strikes obtained from the pair decomposition are sensitive to the BRA output. Small perturbations in the approximate dependence structure can shift the strike locations, which in turn affects the cash component. This sensitivity is more pronounced in higher dimensions because each $Y_i=\sum_{j\ne i}X_j$ aggregates more sources of approximation error. Selecting BRA outputs by $HV$ rather than variance partially mitigates this issue, as documented in the numerical examples.

\section{Conclusion \label{sc:Conclusion}}

This paper provides a simple algorithm for constructing a feasible sub-replicating portfolio for the lower bound of a basket option containing any number of underlying assets. A key methodological contribution is the exact computation of the cash component via vertex enumeration of the piecewise-linear payoff function (Proposition~\ref{prop:exact-lambda}), which removes the Monte-Carlo error that a simulation-based estimate of the cash component would carry. Combined with selecting BRA outputs by portfolio value rather than variance alone, and with a derivative-free strike refinement that directly maximises the hedging portfolio value over the strike locations, this turns the sub-replication framework from a feasibility argument into a verifiable optimality argument: Corollary~\ref{coll:optimality} shows that whenever the hedging-portfolio value matches the basket value under the BRA dependence structure, both are optimal, by weak duality, for the marginals used in the computation. We then carry the argument over to the original marginals. The price of the portfolio under the original marginals is a lower bound, the BRA output reassembled with the original marginals gives an upper bound computed by quadrature, and the two bracket the true lower bound of the basket option (Corollary~\ref{coll:bracket}); in the three- and six-asset log-normal examples the bracket has a relative width of at most $0.1\%$ and $0.02\%$ respectively. A comparison with the multi-marginal optimal-transport algorithm of \cite{neufeld2022mmot} on the same examples shows that the location of the strikes, which the pair decomposition supplies, matters more than their number: our portfolios of two strikes per asset give higher lower bounds than grids of $32$ to $256$ strikes per asset, and on a coarse grid augmented with our strikes no portfolio improves on them. The comparison was run with substitute solvers in place of the commercial solver used by its authors, and their algorithm retains clear advantages in generality and in the number of assets it can handle. The constructed sub-replicating portfolios may serve as a starting point for future work that incorporates additional information on the dependence structure to tighten the bound, and the algorithm could be adapted to other multi-asset derivatives such as spread options. More broadly, this paper contributes a step towards a tractable understanding of lower bounds for basket option prices, and more generally for multi-asset derivatives, in dimensions higher than two.

%

\bibliographystyle{econometrica}
\bibliography{biblio}

@article{hobson2005sub,
  title={Static-arbitrage optimal subreplicating strategies for basket options},
  author={Hobson, David and Laurence, Peter and Wang, Tai-Ho},
  journal={Insurance: Mathematics and Economics},
  volume={37},
  number={3},
  pages={553--572},
  year={2005},
  publisher={Elsevier}
}

@article{laurence2005,
  title={Sharp upper and lower bounds for basket options},
  author={Laurence, Peter and Wang, Tai-Ho},
  journal={Applied Mathematical Finance},
  volume={12},
  number={3},
  pages={253--282},
  year={2005},
  publisher={Taylor \& Francis}
}

@article{pena2010,
  title={Static-arbitrage lower bounds on the prices of basket options via linear programming},
  author={Pe{\~n}a, Javier and Vera, Juan C. and Zuluaga, Luis F.},
  journal={Quantitative Finance},
  volume={10},
  number={8},
  pages={819--827},
  year={2010},
  publisher={Taylor \& Francis}
}

@article{hobson2005sup,
  title={Static-arbitrage upper bounds for the prices of basket options},
  author={Hobson, David and Laurence, Peter and Wang, Tai-Ho},
  journal={Quantitative Finance},
  volume={5},
  number={4},
  pages={329--342},
  year={2005},
  publisher={Taylor \& Francis}
}

@article{neufeld2023,
  title={Model-free bounds for multi-asset options using option-implied information and their exact computation},
  author={Neufeld, Ariel and Papapantoleon, Antonis and Xiang, Qikun},
  journal={Management Science},
  volume={69},
  number={4},
  pages={2051--2068},
  year={2023},
  publisher={INFORMS}
}

@article{neufeld2022mmot,
  title={Numerical method for feasible and approximately optimal solutions of multi-marginal optimal transport beyond discrete measures},
  author={Neufeld, Ariel and Xiang, Qikun},
  journal={arXiv preprint arXiv:2203.01633},
  year={2022}
}

@article{cho2016computing,
  title={Computing lower bounds on basket option prices by discretizing semi-infinite linear programming},
  author={Cho, Hyunseok and Kim, Kyoung-Kuk and Lee, Kyungsik},
  journal={Optimization Letters},
  volume={10},
  number={8},
  pages={1629--1644},
  year={2016},
  publisher={Springer}
}

@article{bernard2018rearrangement,
  title={Rearrangement algorithm and maximum entropy},
  author={Bernard, Carole and Bondarenko, Oleg and Vanduffel, Steven},
  journal={Annals of Operations Research},
  volume={261},
  number={1-2},
  pages={107--134},
  year={2018},
  publisher={Springer}
}

@article{bernard2016algorithms,
  title={Algorithms for finding copulas minimizing convex functions of sums},
  author={Bernard, Carole and McLeish, Don},
  journal={Asia-Pacific Journal of Operational Research},
  volume={33},
  number={5},
  pages={1650040},
  year={2016},
  publisher={World Scientific}
}

@article{boudt2018block,
  title={Block rearranging elements within matrix columns to minimize the variability of the row sums},
  author={Boudt, Kris and Jakobsons, Edgars and Vanduffel, Steven},
  journal={4OR},
  volume={16},
  number={1},
  pages={31--50},
  year={2018},
  publisher={Springer}
}

@article{bernard2026improved,
  title={Improved block rearrangement algorithm},
  author={Bernard, Carole and Chen, Jinghui and R{\"u}schendorf, Ludger and Vanduffel, Steven},
  journal={Annals of Operations Research},
  volume={357},
  number={1},
  pages={605--632},
  year={2026},
  publisher={Springer}
}

@article{puccetti2012computation,
  title={Computation of sharp bounds on the distribution of a function of dependent risks},
  author={Puccetti, Giovanni and R{\"u}schendorf, Ludger},
  journal={Journal of Computational and Applied Mathematics},
  volume={236},
  number={7},
  pages={1833--1840},
  year={2012},
  publisher={Elsevier}
}

@article{bernard2014risk,
  title={Risk aggregation with dependence uncertainty},
  author={Bernard, Carole and Jiang, Xiao and Wang, Ruodu},
  journal={Insurance: Mathematics and Economics},
  volume={54},
  pages={93--108},
  year={2014},
  publisher={Elsevier}
}

@article{wang2016joint,
  title={Joint mixability},
  author={Wang, Bin and Wang, Ruodu},
  journal={Mathematics of Operations Research},
  volume={41},
  number={3},
  pages={808--826},
  year={2016},
  publisher={INFORMS}
}

@article{d2006static,
  title={Static arbitrage bounds on basket option prices},
  author={d'Aspremont, Alexandre and El Ghaoui, Laurent},
  journal={Mathematical Programming},
  volume={106},
  number={3},
  pages={467--489},
  year={2006},
  publisher={Springer}
}

@article{pena2010upper,
  title={Computing general static-arbitrage bounds for European basket options via Dantzig-Wolfe decomposition},
  author={Pe{\~n}a, Javier and Saynac, Xavier and Vera, Juan C. and Zuluaga, Luis F.},
  journal={Algorithmic Operations Research},
  volume={5},
  number={2},
  pages={65--74},
  year={2010}
}

@article{daum2011novel,
  title={A novel feasible discretization method for linear semi-infinite programming applied to basket option pricing},
  author={Daum, Sebastian and Werner, Ralf},
  journal={Optimization},
  volume={60},
  number={10-11},
  pages={1379--1398},
  year={2011},
  publisher={Taylor \& Francis}
}

@article{dhaene1996dependency,
  title={Dependency of risks and stop-loss order},
  author={Dhaene, Jan and Goovaerts, Marc J.},
  journal={ASTIN Bulletin},
  volume={26},
  number={2},
  pages={201--212},
  year={1996},
  publisher={Cambridge University Press}
}

@article{beiglbock2013model,
  title={Model-independent bounds for option prices---a mass transport approach},
  author={Beiglb{\"o}ck, Mathias and Henry-Labord{\`e}re, Pierre and Penkner, Friedrich},
  journal={Finance and Stochastics},
  volume={17},
  number={3},
  pages={477--501},
  year={2013},
  publisher={Springer}
}

@article{aquino2020bounds,
  title={Bounds on multi-asset derivatives via neural networks},
  author={{De Gennaro Aquino}, Luca and Bernard, Carole},
  journal={International Journal of Theoretical and Applied Finance},
  volume={23},
  number={8},
  pages={2050050},
  year={2020},
  publisher={World Scientific}
}

@article{dragazi2024improved,
  title={Improved model-free bounds for multi-asset options using option-implied information and deep learning},
  author={Dragazi, Evangelia and Liu, Shuaiqiang and Papapantoleon, Antonis},
  journal={arXiv preprint arXiv:2404.02343},
  year={2024}
}

@article{eckstein2020robust,
  title={Robust risk aggregation with neural networks},
  author={Eckstein, Stephan and Kupper, Michael and Pohl, Mathias},
  journal={Mathematical Finance},
  volume={30},
  number={4},
  pages={1229--1272},
  year={2020},
  publisher={Wiley Online Library}
}

@article{eckstein2021robust,
  title={Robust pricing and hedging of options on multiple assets and its numerics},
  author={Eckstein, Stephan and Guo, Gaoyue and Lim, Tongseok and Ob{\l}{\'o}j, Jan},
  journal={SIAM Journal on Financial Mathematics},
  volume={12},
  number={1},
  pages={158--188},
  year={2021},
  publisher={SIAM}
}

@article{bartl2022marginal,
  title={Marginal and dependence uncertainty: bounds, optimal transport, and sharpness},
  author={Bartl, Daniel and Kupper, Michael and Lux, Thibaut and Papapantoleon, Antonis and Eckstein, Stephan},
  journal={SIAM Journal on Control and Optimization},
  volume={60},
  number={1},
  pages={410--434},
  year={2022},
  publisher={SIAM}
}

@article{ansari2024improved,
  title={Improved robust price bounds for multi-asset derivatives under market-implied dependence information},
  author={Ansari, Jonathan and L{\"u}tkebohmert, Eva and Neufeld, Ariel and Sester, Julian},
  journal={Finance and Stochastics},
  volume={28},
  number={4},
  pages={911--964},
  year={2024},
  publisher={Springer}
}

@article{kellerer1984duality,
  title={Duality theorems for marginal problems},
  author={Kellerer, Hans G.},
  journal={Zeitschrift f{\"u}r Wahrscheinlichkeitstheorie und verwandte Gebiete},
  volume={67},
  number={4},
  pages={399--432},
  year={1984},
  publisher={Springer}
}

@article{tchen1980inequalities,
  title={Inequalities for distributions with given marginals},
  author={Tchen, Andr{\'e} H.},
  journal={The Annals of Probability},
  volume={8},
  number={4},
  pages={814--827},
  year={1980}
}

@article{ruschendorf1983solution,
  title={Solution of a statistical optimization problem by rearrangement methods},
  author={R{\"u}schendorf, Ludger},
  journal={Metrika},
  volume={30},
  number={1},
  pages={55--61},
  year={1983},
  publisher={Springer}
}

@book{nelsen2006introduction,
  title={An Introduction to Copulas},
  author={Nelsen, Roger B.},
  edition={2},
  series={Springer Series in Statistics},
  publisher={Springer},
  address={New York},
  year={2006}
}

@article{nelder1965simplex,
  title={A simplex method for function minimization},
  author={Nelder, John A. and Mead, Roger},
  journal={The Computer Journal},
  volume={7},
  number={4},
  pages={308--313},
  year={1965}
}

@article{powell1964efficient,
  title={An efficient method for finding the minimum of a function of several variables without calculating derivatives},
  author={Powell, Michael J. D.},
  journal={The Computer Journal},
  volume={7},
  number={2},
  pages={155--162},
  year={1964}
}

@article{ruschendorf2009distributional,
  title={On the distributional transform, {S}klar's theorem, and the empirical copula process},
  author={R{\"u}schendorf, Ludger},
  journal={Journal of Statistical Planning and Inference},
  volume={139},
  number={11},
  pages={3921--3927},
  year={2009},
  publisher={Elsevier}
}

\vspace{3cm}
\newpage
\appendix
\centerline{\Large\bf Appendix}
\section{Proofs} \label{App A}

\subsection{Proof of Proposition~\ref{prop:case1}}

The pointwise inequality $a\le a^+$ for $a\in\mathbb{R}$ gives
$$
x_1+x_2+x_3-K\;\le\;(x_1+x_2+x_3-K)^+,
$$
so $\Pi^{\mathrm{sub}}_1$ is a pointwise sub-replicator, and its expected cost under any $\mu\in\Pi$ is $\mathbb{E}_\mu[X_1+X_2+X_3]-K=m_X-K$, which does not depend on $\mu$. (When $n_1=n_2=n_3=0$ one has $\overline{\nu}_i=\delta_0$, $f_i(z)=z$ and $\overline{\lambda}=\inf_{x\in\mathbb{R}^3_+}\{(\sum_i x_i-K)^+-\sum_i x_i\}=-K$, so the construction of Section~\ref{S1} returns exactly $\Pi^{\mathrm{sub}}_1$.) Taking expectations in the same pointwise inequality shows that $\mathbb{E}_\mu[(X_1+X_2+X_3-K)^+]\ge m_X-K$ for every $\mu\in\Pi$. It therefore remains to exhibit $\mu'\in\Pi$ with $X_1+X_2+X_3\ge K$ $\mu'$-almost surely: for such a $\mu'$ one has $\mathbb{E}_{\mu'}[(X_1+X_2+X_3-K)^+]=m_X-K$, so the lower bound equals $m_X-K$ and $\Pi^{\mathrm{sub}}_1$ attains it. (In particular the hypothesis forces $K\le m_X$.)

Fix an index $i$ with $A_i=\emptyset$, write $F_i$ for the distribution function of $X_i$ and $G$ for that of $Y_i^*=\sum_{j\ne i}X_j^*$ under the dependence structure $\mu^*\in\Pi$ used in the construction, and let $G^{-1}(p):=\inf\{y:G(y)\ge p\}$ and $F_i^{-1}(p):=\inf\{x:F_i(x)\ge p\}$ denote the generalised inverses. No continuity of $F_i$ or $G$ is assumed in what follows. By \eqref{phi1}--\eqref{phi3},
\begin{equation}\label{eq:case1-phi}
\phi_i(x)\;=\;1-\mathbb{Q}(X_i>x)-\mathbb{Q}(Y_i^*\ge K-x)\;=\;F_i(x)-\mathbb{Q}(Y_i^*\ge K-x),
\end{equation}
so $A_i=\emptyset$ means $F_i(x)\le\mathbb{Q}(Y_i^*\ge K-x)$ for all $x\ge 0$.

\noindent\emph{Step 1: the antimonotone coupling of $X_i$ and $Y_i^*$ has sum at least $K$.} Let $U$ be uniform on $(0,1)$ and set $X_i':=F_i^{-1}(U)$, $Y':=G^{-1}(1-U)$. Fix $u\in(0,1)$ and put $x:=F_i^{-1}(u)$, $y:=G^{-1}(1-u)$; by right-continuity of the distribution functions, $F_i(x)\ge u$ and $G(y)\ge 1-u$. Suppose $x+y<K$. Then
$$
\mathbb{Q}(Y_i^*\ge K-x)\;\le\;\mathbb{Q}(Y_i^*>y)\;=\;1-G(y)\;\le\;u\;\le\;F_i(x)\;\le\;\mathbb{Q}(Y_i^*\ge K-x),
$$
where the first inequality uses $y<K-x$ and the last one is $\phi_i(x)\le 0$ (recall that $x\ge 0$). All these quantities are therefore equal, so $\mathbb{Q}(y<Y_i^*<K-x)=0$ and $G(y)=1-u$: the distribution function $G$ is constant, equal to $1-u$, on the non-degenerate interval $[y,K-x)$. A monotone function has at most countably many maximal intervals of constancy, so $1-u$ belongs to a countable set of levels. Hence $\{u\in(0,1):F_i^{-1}(u)+G^{-1}(1-u)<K\}$ is countable, and $X_i'+Y'\ge K$ almost surely.

\noindent\emph{Step 2: embedding into $\Pi$.} Let $(X_j^*)_{j\ne i}$ have their joint law under $\mu^*$, let $Y_i^*=\sum_{j\ne i}X_j^*$, and let $V:=G(Y_i^{*-})+\xi\,\bigl(G(Y_i^*)-G(Y_i^{*-})\bigr)$, where $G(y^-)$ denotes the left limit and $\xi$ is uniform on $(0,1)$ and independent of everything else (the distributional transform of $Y_i^*$, see \cite{ruschendorf2009distributional}). Then $V$ is uniform on $(0,1)$ and $Y_i^*=G^{-1}(V)$ almost surely. Define $X_i:=F_i^{-1}(1-V)$. The vector $(X_1,X_2,X_3)$ obtained in this way has marginals $\mu_1,\mu_2,\mu_3$, i.e.\ its law $\mu'$ belongs to $\Pi$, and $(X_i,Y_i^*)=(F_i^{-1}(1-V),G^{-1}(V))$ is the antimonotone coupling of Step~1 (with $U=1-V$). Hence $X_1+X_2+X_3=X_i+Y_i^*\ge K$ $\mu'$-almost surely, which completes the proof.\qed

\subsection{Proof of Proposition~\ref{prop:case2}}

Throughout write $d_1:=x_1-a_1$, $d_2:=x_2-a_2$ and $d_3:=x_3-c$, so that $x_1+x_2+x_3-K=d_1+d_2+d_3$ because $a_1+a_2+c=K$, and
$$
\Pi^{\mathrm{sub}}_2(x_1,x_2,x_3)=\min(d_1,0)+\min(d_2,0)+\max(d_3,0).
$$

\noindent\emph{(i)} If $d_3\le 0$ then $\Pi^{\mathrm{sub}}_2\le 0\le(d_1+d_2+d_3)^+$. If $d_3>0$ then $\Pi^{\mathrm{sub}}_2=\min(d_1,0)+\min(d_2,0)+d_3\le d_1+d_2+d_3\le(d_1+d_2+d_3)^+$. Hence \eqref{eq:pi2} is a pointwise sub-replicator on $\mathbb{R}^3_+$. With $f_1(z)=\min(z,a_1)$, $f_2(z)=\min(z,a_2)$ and $f_3(z)=(z-c)^+$ one has $\Psi=(x_1+x_2+x_3-K)^+-\Pi^{\mathrm{sub}}_2-(a_1+a_2)\ge-(a_1+a_2)$, with equality at $(x_1,x_2,x_3)=(a_1,a_2,c)$, where both the payoff and $\Pi^{\mathrm{sub}}_2$ vanish; hence $\overline{\lambda}=-(a_1+a_2)$ in \eqref{eq:lambda-def}. The cost follows from $\mathbb{E}[\min(X_i,a_i)]=a_i-P_{X_i}(a_i)$ and $\mathbb{E}[(X_3-c)^+]=C_{X_3}(c)$, and it depends on the marginals only. The bound \eqref{eq:case2-lb} is weak duality (Theorem~\ref{thm:feasible}); the maximum exists because the right-hand side is continuous on the compact simplex $\{a_1+a_2+c=K\}\cap\mathbb{R}^3_+$.

\noindent\emph{(ii)} By continuity of the distribution functions, $\mathbb{Q}(X_i\le a_i)=u^*$ for $i=1,2$ and $\mathbb{Q}(X_3\ge c)=u^*$, and $a_1+a_2+c=S^{\mathrm{c}}(u^*)=K$. Using $\int_0^{u^*}F_{X_i}^{-1}(u)\,du=\mathbb{E}[X_i\1_{\{X_i\le a_i\}}]=a_iu^*-P_{X_i}(a_i)$ for $i=1,2$ and $\int_0^{u^*}F_{X_3}^{-1}(1-u)\,du=\mathbb{E}[X_3\1_{\{X_3\ge c\}}]=C_{X_3}(c)+cu^*$, we obtain
$$
\int_0^{u^*}\bigl(S^{\mathrm{c}}(u)-K\bigr)\,du=(a_1+a_2+c-K)\,u^*+C_{X_3}(c)-P_{X_1}(a_1)-P_{X_2}(a_2),
$$
and the first term vanishes.

\noindent\emph{(iii)} Let $(X_1,X_2,X_3)=(F_{X_1}^{-1}(U),F_{X_2}^{-1}(U),F_{X_3}^{-1}(1-U))$ with $U$ uniform on $(0,1)$; this vector has law $\mu^{\mathrm{c}}$. On $\{U\le u^*\}$ one has $X_1\le a_1$, $X_2\le a_2$ and $X_3\ge c$ by monotonicity of the quantile functions, so $\Pi^{\mathrm{sub}}_2=X_1+X_2+X_3-K=S^{\mathrm{c}}(U)-K=(S^{\mathrm{c}}(U)-K)^+$, the last equality because $S^{\mathrm{c}}\ge K$ on $(0,u^*]$. On $\{U\ge u^*\}$ one has $X_1\ge a_1$, $X_2\ge a_2$ and $X_3\le c$, so $\Pi^{\mathrm{sub}}_2=0=(S^{\mathrm{c}}(U)-K)^+$ because $S^{\mathrm{c}}\le K$ on $[u^*,1)$. Hence $\Pi^{\mathrm{sub}}_2=(X_1+X_2+X_3-K)^+$ $\mu^{\mathrm{c}}$-almost surely and, taking expectations, $C_{X_3}(c)-P_{X_1}(a_1)-P_{X_2}(a_2)=V(\mu^{\mathrm{c}})$. Writing $v:=C_{X_3}(c)-P_{X_1}(a_1)-P_{X_2}(a_2)$ and combining this with (i) and with $\mu^{\mathrm{c}}\in\Pi$,
$$
v\;\le\;\inf_{\mu\in\Pi}\mathbb{E}_\mu\bigl[(X_1+X_2+X_3-K)^+\bigr]\;\le\;V(\mu^{\mathrm{c}})=v,
$$
so the three quantities coincide: $\mu^{\mathrm{c}}$ is optimal and \eqref{eq:pi2} is an optimal sub-replicating portfolio. The expression as $\int_0^{u^*}(S^{\mathrm{c}}(u)-K)\,du$ is (ii), and under the single-crossing hypothesis this integral equals $\int_0^1(S^{\mathrm{c}}(u)-K)^+du=V(\mu^{\mathrm{c}})$.\qed

\subsection{Proof of Proposition~\ref{prop:case3}}

Let $\mu^*\in\Pi$ be a dependence structure exhibiting joint mixability, so that $X_1+X_2+X_3$ is $\mu^*$-almost surely equal to a constant $c$. Taking expectations gives $c=\mathbb{E}_{\mu^*}[X_1+X_2+X_3]=m_X$, since marginal-mean preservation $\mathbb{E}_{\mu}[X_i]=\mathbb{E}[X_i]$ holds for every $\mu\in\Pi$. Hence the constant is necessarily $m_X$. By Jensen's inequality applied to the convex function $z\mapsto(z-K)^+$, for any $\mu\in\Pi$,
$$
\mathbb{E}_\mu[(X_1+X_2+X_3-K)^+]\;\ge\;(\mathbb{E}_\mu[X_1+X_2+X_3]-K)^+\;=\;(m_X-K)^+.
$$
Equality holds at $\mu=\mu^*$ because $X_1+X_2+X_3-K$ is the constant $m_X-K$ under $\mu^*$ and hence equals its positive part when $K\le m_X$, or equals zero almost surely when $K>m_X$. Therefore the lower bound equals $(m_X-K)^+$.

\medskip

\noindent\textit{Case $K\le m_X$.} Under $\mu^*$, $x_1+x_2+x_3-K=m_X-K\ge 0$, so $(x_1+x_2+x_3-K)^+=m_X-K$. The pointwise inequality $a\le a^+$ gives
$$
x_1+x_2+x_3-K\;\le\;(x_1+x_2+x_3-K)^+\quad\forall\,(x_1,x_2,x_3)\in\mathbb{R}_+^3,
$$
so $\Pi^{\mathrm{sub}}_1(x_1,x_2,x_3)=x_1+x_2+x_3-K$ is a pointwise sub-replicator. Its expected cost under any $\mu\in\Pi$ is $m_X-K$, which equals the lower bound. Hence $\Pi^{\mathrm{sub}}_1$ is optimal.

\medskip

\noindent\textit{Case $K>m_X$.} Under $\mu^*$, $x_1+x_2+x_3=m_X<K$ almost surely, so $(x_1+x_2+x_3-K)^+=0$ $\mu^*$-almost surely and the lower bound is $0$. The trivial zero portfolio sub-replicates pointwise (since $0\le (x_1+x_2+x_3-K)^+$ for all non-negative $x_i$) and its expected cost equals $0$, attaining the lower bound. In this regime the zero-crossings of the functions $\phi_i$ do not lead to the zero portfolio: writing $e_i$ and $s_i$ for the essential infimum and supremum of $X_i$ and assuming $F_{X_i}$ continuous and strictly increasing on $[e_i,s_i]$, one finds from \eqref{eq:case1-phi} with $Y_i^*=m_X-X_i^*$ that $\phi_i(x)=F_{X_i}(x)-F_{X_i}(x-K+m_X)$, so $A_i=(e_i,\,s_i+K-m_X)$ and $f_i(z)=\min(z,e_i)+(z-s_i-K+m_X)^+$. With the convention of Algorithm~\ref{alg:subrep}, which moves the last strike to $K$ because it exceeds the largest atom $s_i$ of $X_i$, one obtains $\mathbb{E}[f_i(X_i)]=e_i$ and $\overline{\lambda}=-\sum_i e_i$ (the bound $\Psi\ge-\sum_i e_i$ follows from $\sum_i(x_i-K)^+\le(\sum_i x_i-K)^+$, with equality at $x=(e_1,e_2,e_3)$), hence a portfolio of expected cost $0$, which is optimal as well; without that convention the expected cost can be strictly negative. In practice the basket is only approximately constant under the BRA output and the functions $\phi_i$ are then dominated by numerical noise, which is why the implementation evaluates the portfolios without strikes first (Remark~\ref{rem:trivial}): the portfolio $\sum_i(x_i-K)^+$ has expected cost $\sum_iC_{X_i}(K)=0$, since $X_i\le m_X<K$ almost surely, and is returned without any strike search.\qed

\end{document}